\documentclass[final,5p,times,twocolumn]{elsarticle}
\usepackage{amssymb}
\usepackage{amsthm}
\begin{document}

\begin{frontmatter}



\title{Self-assembly of magnetic iron oxide nanoparticles into cuboidal superstructures }


\author[L1]{Sabine Rosenfeldt} 
\author[L2]{Stephan F\"{o}rster}
\author[L3]{Thomas Friedrich} 
\author[L4]{Ingo Rehberg}
\author[L5]{Birgit Weber}

\address[L1]{Physical Chemistry I, University of Bayreuth}
\address[L2]{ICS-1, Forschungszentrum J\"{u}lich}
\address[L3]{Institute of Medical Engineering, University of L\"{u}beck}
\address[L4]{Experimentalphysik V, University of Bayreuth}
\address[L5]{Anorganische Chemie II, University of Bayreuth}
\begin{abstract}
This chapter describes the synthesis and some characteristics of magnetic iron oxide nanoparticles, mainly nanocubes, and focus on their self-assembly into crystalline cuboids in dispersion. The influence of external magnetic fields, the concentration of particles, and the temperature on the assembly process is experimentally investigated.   
\end{abstract}


\end{frontmatter}

\tableofcontents 
%\newpage						%nur zum Platzsparen rauskommentiert! 10.11.17

\newcommand{\figwidth}{9cm} %maximal 8cm
%\input{T3_introduction}%T3
%spell-checker britisch 10.11.17, 17 uhr, Ingo
\section{Motivation}

\subsection{Properties of magnetic iron oxide nanoparticles}

Nanoparticles possess at least in one dimension a size in the nanometer range. Their properties can differ from the ones of the corresponding bulk material.  Spherical particles with a radius of 10\,nm consist of about 8000\,atoms, in which  6\,\% are surface atoms. In contrast, a sphere with a radius of 1\,mm exhibits only  $ 6 \cdot 10^{-5}$\,\% surface atoms. In consequence of the energetically unfavourable ratio of surface to volume nanoparticles tend to aggregate. Colloidal stability of nanoparticles can be achieved by a protective cover. In case of short-range attractive interaction the protecting layer screens attractive interaction such that the nanoparticles form stable dispersions. Hence, the resulting nanoparticles are by default surface-functionalized and often show a core-shell structure. 

Nanoparticles exhibit physical properties that depend strongly on their size and shape. Magnetic nanoparticles may possess one or more than one Weiss-domain. Depending on their size and the temperature they show ferromagnetic or superparamagnetic behaviour. If the nanoparticles are small enough they contain only one magnetic domain. At room temperature most sub-10\,nm iron oxide magnetic nanoparticles show superparamagnetic behaviour because the thermal energy is sufficient to flip the magnetization direction. The temperature dependence of the fluctuation time $\tau_\mathrm{N}$ of the magnetization has been described by N\'{e}el as
$
\tau_{\mathrm{N}} = \tau_0 \exp \left( \frac{\Delta E_\mathrm{mag}}{k_\mathrm{B}T}\right) 
$,  where the $\Delta E_\mathrm{mag}$ is the energetic barrier between the two magnetization states, $k_\mathrm{B}$ the Boltzmann constant and $T$ the temperature. $\tau_0$ is a characteristic time in the order of nanoseconds.  
Below the so called blocking temperature, the flipping of the magnetization is extremely rare and hysteresis is observed, i.e., the material is ferromagnetic. As $\Delta E_\mathrm{mag}$ depends strongly on size, anisotropy and  surface properties of the nanoparticles, so does the blocking temperature. 

%\input{T4_applications}%T4
%spell-checker britisch 10.11.17, 12 uhr, Ingo
\subsection{Applications of magnetic iron oxide nanoparticles}

Nanostructured magnetic materials are omnipresent. They are used in medical, environmental and technical applications.\,\cite{JMaterChem-2011-21-16819-16845}  Iron oxide nanoparticles consist mostly of maghemite ($\mathrm{\gamma -Fe_2O_3}$) or magnetite ($\mathrm{Fe_3O_4}$) and exhibit single domains of about 5-20\,nm in diameter. \cite{SPION-groesse} Magnetite is a cubic inverse spinel. The oxygen form a face centred cubic packing and the iron cations occupy interstitial tetrahedral (tetr) and octahedral (octa) sites. The formula can be written as $\mathrm{(Fe^{3+})^{tetr} (Fe^{3+}, Fe^{2+})^{octa} O_4}$.
The electrons can hop between $\mathrm{Fe^{2+}}$  and $\mathrm{Fe^{3+}}$ ions in the octahedral sites at room temperature.
Maghemite can be considered as an $\mathrm{Fe^{2+}}$-free  magnetite. It results the formula
$\mathrm{\left( Fe^{3+}_8 \right) ^{tetr} \left( Fe^{3+}_{40/3} o _{8/3} \right) ^{octa} O_{32}}$
(vacancies $o$). \cite{cornell-SPION-zusammensetzung}

For medical and bioengineering in vivo applications superparamagnetic iron oxide nanoparticles (SPIONs) with high magnetization and sizes smaller than 100\,nm are commonly used, because these nanoparticles are non-toxic and biocompatible. \cite{SPIONs-ChemRev2008-108-2064} 
The medical applications make use of the magnetic moment of the particles in three different ways: (i) For  hyperthermical applications parts of the in vivo-material are heated locally by external time-dependent magnetic field. %lit
(ii) For magnetic particle imaging the non-linear response of the magnetization is locally detected and interpreted as an indication for the particle concentration. \cite{buch2} (iii) Surface functionalized magnetic nanoparticles are used for targeted drug delivery, where SPIONs are addressed by external magnetic fields and located in a specific area of the body.  \cite{SPIONs-ChemRev2008-108-2064}

Iron oxide nanoparticles are also used in environmental application.\,\cite{Nanoscale-Research-Letters-2016-11-498} Suitably coated with catalysts or enzymes, these nanoparticles  can be used as cleaning agent, which can be extracted by external magnetic fields. In addition these particles can be easily detected. Such strategies are used in oil-field rocks and contaminated geological systems.\,\cite{EnvironSciTechnol-2014-48-1189} 

Many technical applications are based on ferrofluids, which consist of magnetic nanoparticles in a non-magnetic solvent. Examples are damping devices or magnetic seals for rotating shafts.  \cite{Rosensweig} The field and temperature dependent variation in the refractive index of these fluids is interesting for optical filters, optical gratings or defect sensors. \cite{Nanofluids-3-1-2012} \cite{Nanofluids-5-1-2016}

%\input{T5_saxs}%T5 %Fig 1-3
%spell-checker britisch 10.11.17, 17 uhr, Ingo

\section{Synthesis and properties of nanoparticles}

%\begin{enumerate}
%\item Synthese allgemein
%\item kurze Worte Charakterisierung allgemein
%\item Inhalt Kapitel
%\end{enumerate}

Synthesis of well defined biocompatible, monodisperse iron oxide nanoparticles in aqueous media is demanding for three reasons: (i)  Generally, the synthesis is done in organic media and a solvent transfer is performed later on. (ii) Surface functionalization of the nanoparticles is crucial to circumvent  aggregation and surface-oxidation. (iii) For monodispersity the reaction parameters and time must be precisely controlled.

Common methods to synthesize iron oxide nanoparticles are (i)co-pre\-ci\-pi\-ta\-tion, (ii) thermal decomposition, and (iii) hydrothermal or solvothermal  synthesis.  Moreover, the trend to green-chemistry leads to the development of (iv) biological synthesis routes. \cite{SciTechnolAdvMater-16-2015-02350}
(i) During co-precipitation ferric ($\mathrm{Fe^{3+}}$) and ferrous ($\mathrm{Fe^{2+}}$) ions are mixed in basic solution. It's a classical way to obtain large amounts of iron oxide nanoparticles with a high saturation magnetization, but with the disadvantage of a broad particle size distribution.
(ii) Monodisperse and highly crystalline particles are obtained via high-temperature thermal decomposition. The size of the nanoparticles is adjustable by parameters like the ageing temperature and the\begin{scriptsize}
{\footnotesize â€¢}
\end{scriptsize} salt concentration. (iii) Hydrothermal and solvothermal  syntheses  are wet-chemical techniques of crystallization under high pressure. The synthesis routes allow a good control over the chemical composition and the shape of the iron oxide nanoparticles. This way, even capsules, nanotubes or  hollow iron oxide nanoparticles can be produced. Such an architectural control requires a heroic synthetic effort. (iv) Biological approaches use enzymes or bacteria for the reduction of salts and the conversion into the respective iron oxide nanoparticles. For example, actinobacter spp.\ reacts with ferric chloride precursors to maghemite under aerobic conditions. Biosynthesis is eco-friendly, but the morphological control of the final nanoparticles is still in its infancy.

\subsection{Synthesis of spherical and cubic iron oxide nanoparticles } \label{synthesis}

The iron oxide particles, which are discussed in the following, are synthesized by the thermal decomposition method originally invented by Park et al.\ (2004).\,\cite{Nature-materials-3-891-2004}
Following this approach, small dispersed iron oxide nanoparticles are produced in a large-scale synthesis. Typically,  iron oxide nanoparticles  with dimensions between 5-30\,nm are obtained. They are monodisperse due to a separation of nucleation and growth.
The synthetic procedure bases on the thermal decomposition of iron oleate precursors in a high boiling solvent, like 1-hexadecene (b.p.\,274\,$^{\circ}$C) or 1-octadecene (b.p.\,317\,$^{\circ}$C). The reactivity of the iron oleate complex and thereby, the size of the nanoparticle, increases with increasing boiling point of the solvent. The metal-oleate precursor is prepared from reacting an iron salt and sodium oleate. The morphology of the iron oxide nanocrystals can be finer adjusted by the oleic acid concentration. Spherical nanoparticles are obtained with small excess of sodium-oleate, whereas cubic particles results for larger excess of stabilizing ligand. The transition from spherical to cubic occurs around an sodium oleate excess of ca. 5\,\% in the precursor complex.
The specific synthesis of iron oxide nanoparticles, discussed in this chapter, is summarized in table 1.
%\ref{rezept}.
%-hier urspruenglich Tabelle \label{rezept}
\begin{table} \label{rezept}
\begin{center}
\begin{tabular}{p{8cm}}
\hline
\textbf{Iron(III)-oleate precursor}\\
\hline
21.60 g $\mathrm{FeCl_3 \cdot 6 \, H_2O}$, 73.00-80.3\,g sodium oleate (120-1200\,mmol, corresponding to 0-10\,\% excess), 160\,mL ethanol, 120\,ml distilled water and 180\,mL hexane  are suspended and heated to 70$\,^{\circ}$C for 4\,h.  After cooling to room temperature, the iron oleate complex containing organic phase is washed with water several times. After purification the solution is concentrated under vacuum conditions  at 110$ \,^{\circ}$C until the red-brownish iron-oleate complex shows up in a waxy form.  \\
\hline
\textbf{Iron oxide nanoparticles}\\
\hline
Typically, 36\,g (40\,mmol) of the iron oleate complex were dissolved in a mixture of 200\,g 1-octadecene and 2.8-5.7\,g oleic acid. The spherical particle shown in Fig.\,2
%\ref{idee-birgit-HRTEM} 
is prepared with 5.7\,g oleic acid and 0\,\% sodium oleate excess during the synthesis of the precursor, and the cubic one with 2.85\,g oleic acid and 10\,\% sodium oleate excess in the iron oleate complex synthesis. The reaction mixture was heated up under stirring to 110$\,^{\circ}$C with a heating rate of  ca. 2\,K/min. After insertion of nitrogen cover gas, the mixture is heated further to 318\,$^{\circ}$C and kept at this temperature for about 15\,min. Thereby the colour of the solution turns to black. It indicates the formation of the iron oxide nanocrystals. At room temperature the volume is doubled with tetrahydrofurane. Precipitation of the nanoparticles is performed by adding acetone.  The nanocrystals were separated by centrifugation (24\,h with 4800\,rpm, ca.\,2500\,$\mathrm{g}$). After decant of the upper phase, the iron oxide nanoparticles can be redissolved in organic solvents like toluene. \\
\hline
\end{tabular}
\caption{Synthesis of iron oxide nanoparticles (cf. Fig.\,2
%\ref{idee-birgit-HRTEM}) 
and the basic precursors. The ratio of iron chloride to sodium oleate determines the shape of the iron oxide nanoparticles later on.}
\end{center}
\end{table}

%--zusammensetzung nanocrystals BW
%
\begin{figure} \label{idee-birgit-moessbauer}
	\centering
	\includegraphics[width=\figwidth]{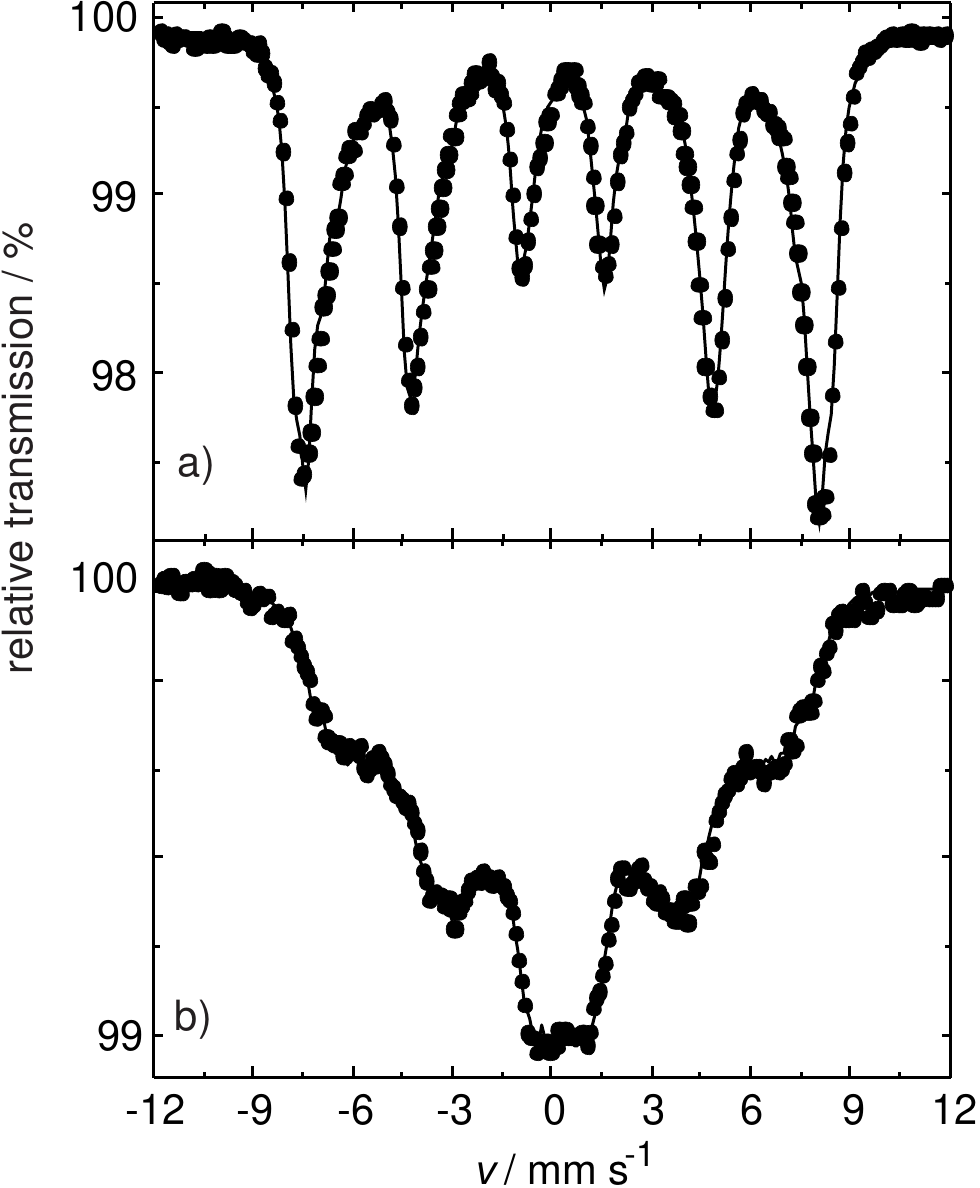}
	\caption{M\"{o}ssbauer spectra of iron oxide nanoparticles at 77\,K. Both particles are stabilized by oleic acid. a) Iron oxide cubes with a diameter of 9\,nm. The composition is 60\,\% maghemite and 40\,\% magnetite. b) Iron oxide spheres of 5\,nm size. }
\end{figure}
The composition of the nanocrystals is $\mathrm{ \left(\gamma -Fe_2O_3 \right) _ {1-x} \left( Fe_3O_4 \right) _ {x}} $ ($\mathrm{x} \in [0,1] $).\,\cite{Nature-materials-3-891-2004} The oxidation of magnetite to maghemite leads to a lower magnetization of the nanoparticle. Park et al.\ report that magnetite dominates in small 5\,nm sized spherical nanocrystals ($\mathrm{x=0.8}$), while the amount of maghemite increases with the particle size  ($\mathrm{x=1}$ for a size of about 20\,nm). The ratio of the two iron oxides in the nanocrystals can be determined by
$^{57}\mathrm{Fe}$-M\"{o}ssbauer spectroscopy. This technique bases on a recoilless absorption of high energy $\gamma$-quanta, which lead to small changes in the energy levels of atomic nuclei. For iron species the electron transitions from nuclei with orbital quantum moment of $I= 3/2$ to $I= 1/2$ are detected.  Two examples of such $^{57}\mathrm{Fe}$-M\"{o}ssbauer spectra are shown in Fig.\,1. 
% \ref{idee-birgit-moessbauer}. 
The characteristic sextets of $\mathrm{Fe^{3+/2+}}$-ions in tetrahedral and octahedral sites are visible.

A quantitative analysis of the spectrum shown on the top reveals 60\,\% maghemite and 40\,\% magnetite. The spectrum at the bottom is of nanocrystals with a core size of 5\,nm. The line width increases with decreasing particle size, prevailing a determination of the exact composition in this case. \cite{Nature-materials-3-891-2004}

%-----------------
\begin{figure} \label{idee-birgit-HRTEM}
	\centering
	\includegraphics[width=\figwidth]{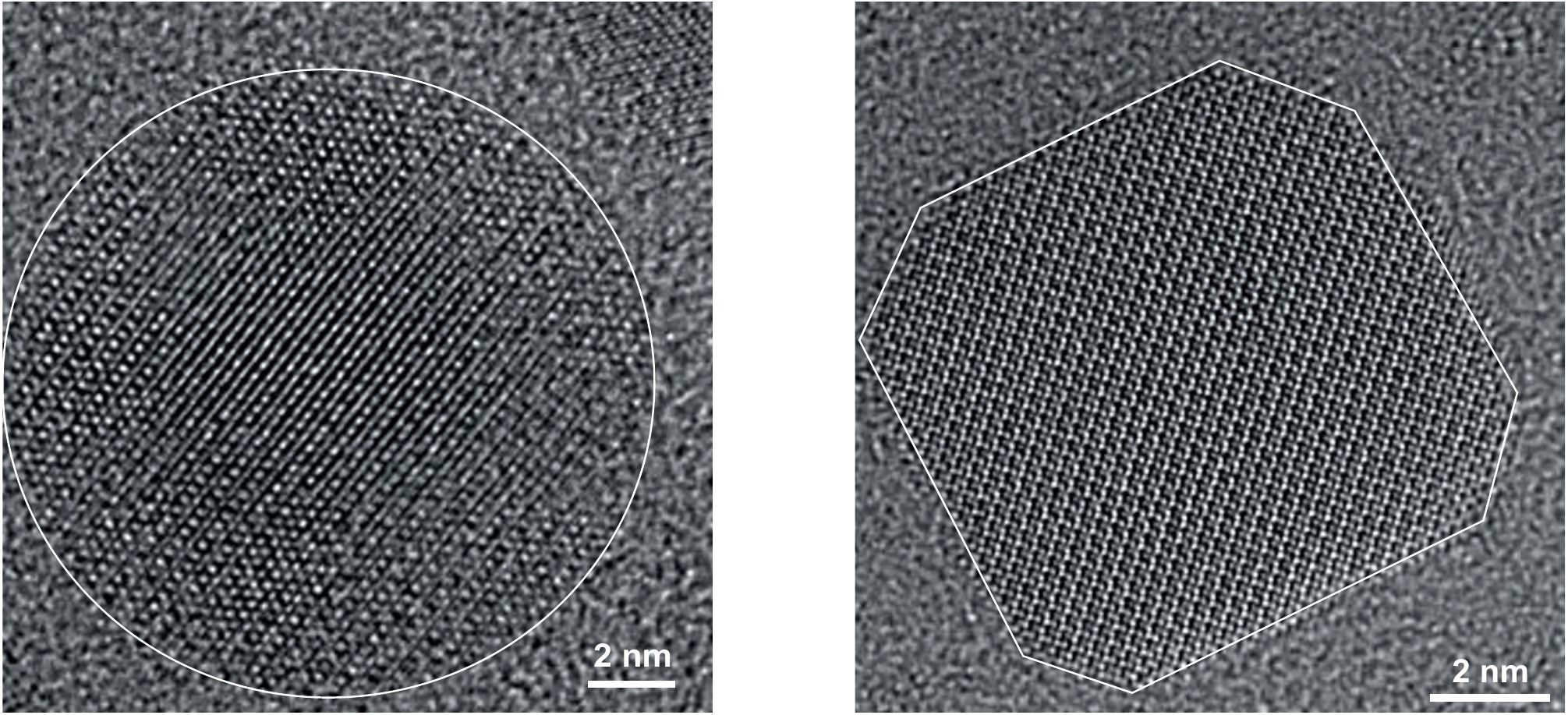}
	\caption{STEM images of spherical (left) and cubic (right) nanoparticles, which consist of a crystalline  iron oxide core (nanocrystal) and an oleic acid shell. The brightness of the images are determined by the electron density, thus the method is sensitive to the iron oxide core of the nanoparticles. The size of the nanocrystals are 13\,nm (left) and 9\,nm (right). The oleic acid shell contrast is to low to be detectable.	}
\end{figure}

The shape and size of the nanoparticles is further analysed by electron microscopy. Figure 2
%\ref{idee-birgit-HRTEM}
shows high resolution transmission microscopy images (STEM) of cubic (right) and  spherical (left) nanoparticles, which are synthesised using a precursor wax with 10\,\% and 0\,\% sodium oleate excess. Due to the high electron contrast of iron species only the crystalline iron oxide core (the nanocrystal) is visible in STEM, whereas the oleic acid shell is invisible. The images show the nearly perfect geometry of the iron oxide nanocrystals. The spherical shape is nearly perfect, and even the cubic nanocrystal is only slightly truncated.
A 2D-Fourier analysis of such images, followed by radial averaging (not shown here), reveals that the crystal structure  is compatible with an inverse spinell type.

Complementary information about the shape and size of the particles can be obtained by small-angle x-ray scattering (SAXS), where the particles are dispersed in a fluid. Two examples are given in the insets of Fig.\,3.%\ref{spheres-cubes-characterization}. 
The x-ray scattering is mainly determined by the metal oxide core of the particles, whereas the stabilizing shell is barly detectable. Quantitatively speaking, the excess electron density for the core is about 1200\,$\mathrm{nm}^{-3}$, which is about 20\,times larger than the excess electron density of the oleic acid shell.

\begin{figure} \label{spheres-cubes-characterization}
	\centering
	\includegraphics[width=\figwidth]{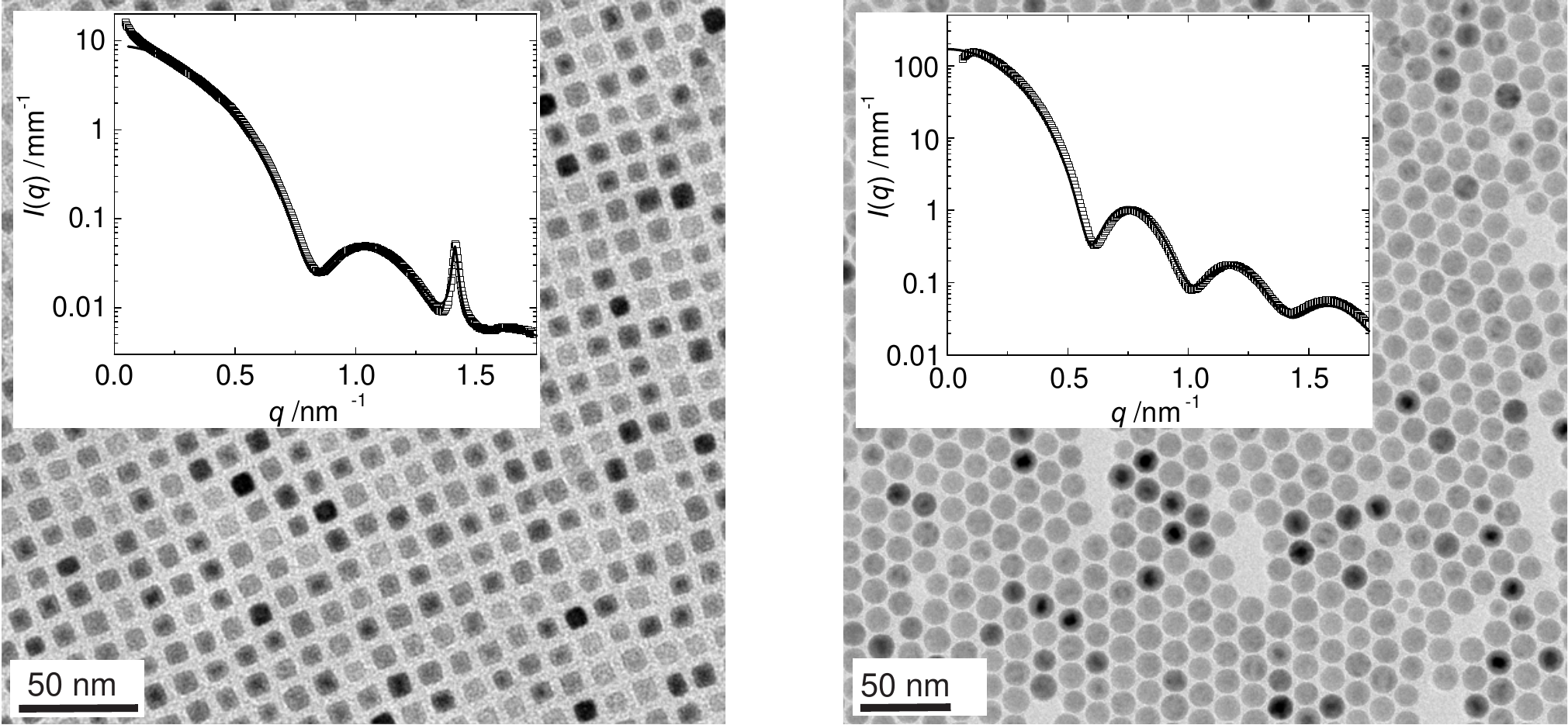}
	\caption{Cyro-TEM (0.1\,wt\%) and SAXS data  (2\,wt\%, insets) of iron oxide nanoparticles, which are stabilized by an oleic acid layer. Due to the low contrast (excess electron density) of the stabilizing oleic acid layer their x-ray scattering behaviour is mainly determined by the metal-oxide core of the particles (excess electron density 1470\,$\mathrm{nm^{-3}}$ for iron oxide, 300\,$\mathrm{nm^{-3}}$ for oleic acid and  238\,$\mathrm{nm^{-3}}$ for toluene).
The fits (lines in insets) reveal  single cubes (9\,nm) and spheres (13\,nm). The Bragg reflex at $q = 1.41\,\mathrm{nm^{-1}}$ seen in the SAXS pattern on the left side is attributed to crystalline oleic acid. The conditions under which this sharp peak develops are still unclear.}
\end{figure}

The SAXS intensity $I(q)$ is proportional to the product of the  form factor $P(q)$ and the structure factor $S(q)$. $P(q)$ is determined by the shape, and $S(q)$ refers to the arrangement of the particles. At low concentrations the scattering is determined by the particle shape, e.\,g.\ mainly by the form factor scattering of the iron oxide nanoparticles. The form factor of a homogeneous cube differs marginally from the one of a homogeneous sphere --- the wavelength of the oscillations is broader and the deepness of the minima is smaller for cubes compared to the ones of volume identical spheres. The intensity decays in both cases with $q^{-4}$. The experimental data (open squares) can well be described by the corresponding form factor $P(q)$ indicated by the solid lines, when allowing for a small polydispersity. The fit of $P(q)$ leads to cubes of 9\,nm and spheres of 13\,nm. The standard derivation results in both cases to 1\,nm, when assuming a Gaussian particle distribution in diameter.

The measured intensity of the cubic nanoparticles have an additional feature, namely a sharp peak at $q = 1.41\,\mathrm{nm^{-1}}$. This scattering vector corresponds to a crystalline structure with a characteristic length of 4.5\,nm in real space. This reflex is attributed to oleic acid, as shown in Fig.\,3
%\ref{cubes2charcterization-cryoTEM} 
and discussed in detail later. Deviations between the fit and the data at scattering vectors smaller than $q < 0.15 \,\mathrm{nm^{-1}}$ are an indication for a minor fraction of aggregates, as described by Klokkenburg et al. (2007) \cite{PhysRevE-2007-75-051408} for magnetite ferrofluids.

The size of the iron oxide core obtained by SAXS is significantly smaller than the  hydrodynamic radius obtained by dynamic light scattering (DLS).  In DLS the overall size of the nanoparticle is seen, which includes the stabilizing oleic acid layer. The corresponding hydrodynamic diameters of the particles are 19.5\,nm for the spherical and 16.7\,nm for the cubic nanoparticles. The monodispersity is expressed by a polydispersity index of PDI $=$ 0.04 for the spheres and PDI $=$ 0.08 for the cubes.  \cite{DLSbuch} The hydrodynamic radius obtained by DLS measurements leads to a oleic acid layer thickness of about 7\,nm for one nanocube.

The core-shell morphology of  oleic acid stabilized iron oxide nanoparticles is also clearly visible in the cryo-TEM images. The cubes show an almost perfectly ordered lattice. The oleic acid layer thickness can be estimated from the lattice plane distance to be about 5\,nm.  The peak seen at $q = 1.41\,\mathrm{nm}^{-1}$ in the SAXS pattern corresponds to this distance. The smaller value of the oleic acid shell obtained from cryo-TEM analysis is explained by the fact that the particles overlap within their hydrodynamic radii in the packed state, which indicates a denser packing of oleic acid in the latter one.

\subsection{Magnetic properties of the nanoparticles}

For sufficiently small single domain ferromagetic or ferrimagnetic nanoparticles the particles become superparamagnetic. For superparamagnetic na\-no\-par\-tic\-les, in the absence of an external magnetic field, the time necessary to measure the magnetization of the nanoparticles is significantly longer than the N\'{e}el relaxation time $\tau_\mathrm{N}$ and therefore the magnetization appears to be zero. In the presence of an external magnetic field, the nanoparticles are magnetized, however, the magnetic susceptibility is much larger compared to a paramagnet. The measurement time %($\tau_\mathrm{m}$) 
strongly depends on the method used. If the magnetic measurement time $\tau_\mathrm{m}$ is much larger than $\tau_\mathrm{N}$, the magnetization flips many times during the measurement and the measured magnetization averages to zero. If $\tau_\mathrm{m}$ is much smaller than $\tau_\mathrm{N}$, the nanoparticle shows a magnetic moment. The crossover from the ferromagnetic to the superparamagnetic state is observed when $\tau_\mathrm{m}$ and $\tau_\mathrm{N}$ have the same order of magnitude. Operationally, this transition is detected as a maximum in the temperature dependent susceptibility. The corresponding temperature is called blocking temperature.

\begin{figure} \label{idee-birgit-2}
	\centering
	\includegraphics[width=\figwidth]{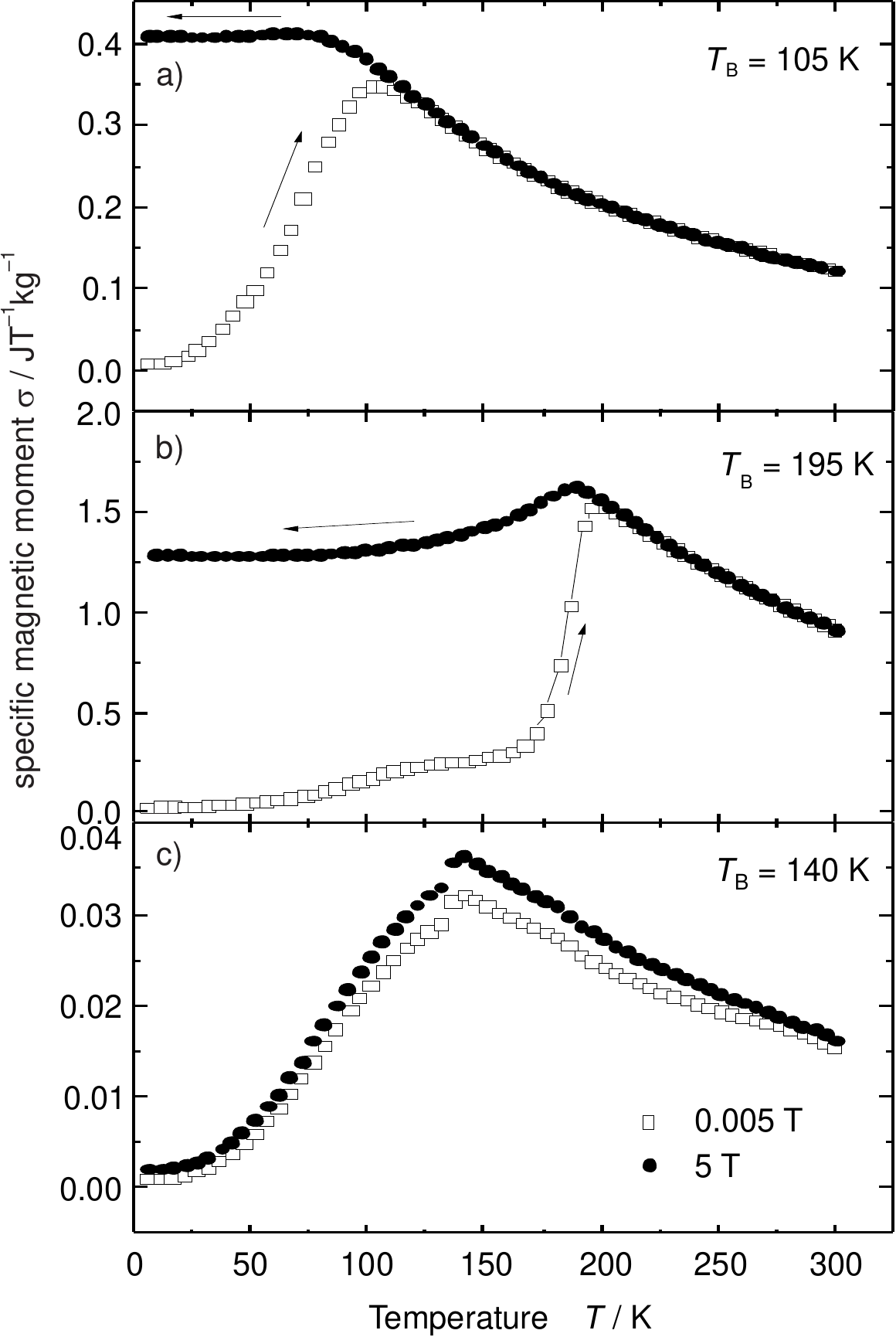}
	\caption{FCM and ZFCM measurements of a) dried cube-shaped nanoparticles (9\,nm), b) dried spherical nanoparticles (13\,nm) and c) a suspension of cube-shaped nanoparticles (9\,nm) in toluene/THF.}
\end{figure}

In Fig.\,4
%\ref{idee-birgit-2}, 
the temperature dependent specific magnetic moment $\sigma$ of cube shaped and spherical nanoparticles is given. In Fig.\,4\,a,
%\ref{idee-birgit-2}\,a, 
both, zero field cooled measurement (ZFCM) and field cooled measurement (FCM) data of dried cube shaped nanoparticles with an average size of 9 nm are displayed. The ZFCM data are shown as open squares and the FCM data as filled circles. For the FCM measurements, the sample is cooled down while measuring the magnetization using a small external magnetizing field (0.005\,T). The specific magnetic moment is detected using a vibrating sample. It decreases monotonically with increasing temperature. In preparation for the ZFCM measurements, the external magnetizing field was set to zero at room temperature and the sample was cooled down to 2\,K. Then a magnetizing field of 0.005\,T was turned on and the magnetization is measured for increasing temperature. The specific magnetic moment increases up to 105\,K and then decreases upon further heating. This maximum corresponds to the blocking temperature.

In Fig.\,4\,b,
%\ref{idee-birgit-2}\,b, 
the same measurement is performed for dried spherical nanoparticles with an average size of 13\,nm. The FCM measurements do not show a monotonic behavior in this case, a fact that we cannot explain. The blocking temperature derived from the ZFCM measurements is located at 195\,K. This is about a factor of two larger and reflects the larger volume of this particles.

In Fig.\,4\,c,
%\ref{idee-birgit-2}\,c, 
the ZFCM measurements of cubic nanoparticles in suspension are presented. In preparation of the ZFCM measurements, the nanoparticles in a toluene and tetrahydrofurane (THF) solvent mixture were exposed to an external magnetic field at room temperature for 30\,min. The motivation behind this protocol was to trigger field-induced self assembly of the nanoparticles. Please note the melting points of the solvents toluene (178\,K) and tetrahydrofurane (165\,K) are in the neighbourhood of the maximum seen at 140\,K. Thus this maximum might partly be caused by a liquid to solid phase transition, so that an interpretation of this curve seems to be difficult at this time. With increasing external magnetic field a slight increase of $\sigma$ is observed. This could be an indication for an increase in particle size through self assembly.

In Fig.\,5
%\ref{idee-birgit-Hys} 
the field dependence of $\sigma$ at 300\,K and 10\,K is shown. At 300\,K (open squares) in the superparamagnetic state, the magnetization curve is a reversible S-shaped increasing function. The increasing applied field leads to an increasing alignment of the magnetic moments of the superparamagnetic nanoparticles along the applied field. At 10\,K (filled circles) the nanoparticles are in the blocked state and a hysteresis of the magnetization is observed as expected for a ferro- or ferrimagnetic material.

\begin{figure} \label{idee-birgit-Hys}
	\centering
	\includegraphics[width=\figwidth]{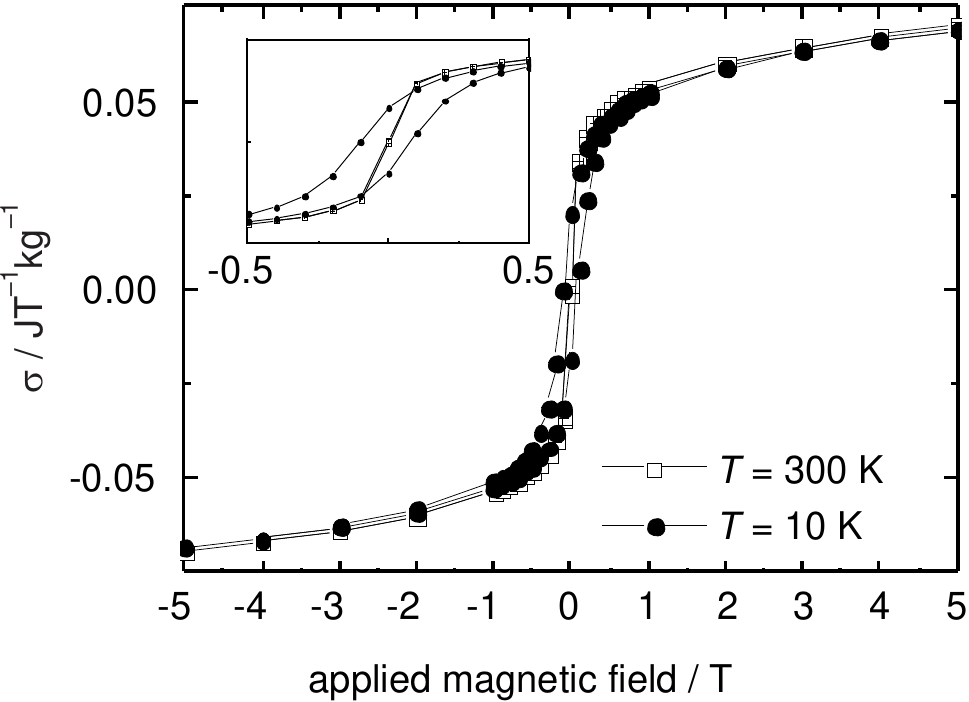}
	\caption{Field dependence of $\sigma$ of dried cube-shaped nanoparticles (9\,nm) at 300\,K and 10\,K. The inset shows a magnification of the hysteresis range.}
\end{figure}

\section{Self-assembly into superstructures} \label{self-assembly}
%\input{T7_assembly}%T7 
%--------T7: SAXS in concentrated solution

One of the most fascinating processes in nature is self-assembly. In the context of this chapter the term self-assembly is used in the following way: A transition from isolated nanoparticles dispersed in a fluid to a stable crystal-like assembly. This can lead to a huge amount of structures, including chains, helices, sheets, cylinders, well-defined 3D-superlattices, gyroids or disordered clusters. 

For magnetic nanoparticles, van der Waals attraction and effective dipole-dipole interaction of magnetic nanoparticles can lead to well-ordered assembly structures.\,\cite{Hindawi} \cite{frontiers-in-china} \cite{SciTechnolAdvMater-15-20142014-055010} 
The assembly of monodisperse magnetic nanoparticles into superstructures is considered as an important step toward fabrication of functional devices, like catalysts, %\textbf{lit cat,target-storage},
targeting or storage media. With ongoing miniaturization of these applications, a lower size limit for the magnetic nanoparticles may be reached. In this limit the contributions of magnetic interaction energy, thermal energy and other interaction energies, such as van-der-Waals and interfacial interaction energy, may be comparable. 

In this section the self-assembly of iron oxide nanoparticles is addressed.
Faceted and, in particular, cubic nanoparticles forming lattice matched superstructures are of intriguing interest for materials with high magnetic anisotropy constant because of the possibility to align the easy magnetization axis of individual nanoparticles, leading to a new kind of artificial magnetic solid. Two and three dimensional self-assembled structures with controlled micro- and meso-scaled ordering consisting of iron oxide nanocubes can be produced by solvent-evaporation within  a magnetic field. \cite{NanoLett-2011-11-1651} \cite{PNAS-2007-6-17570} \cite{Nanoscale-2016-8-15571} Thereby the mesocrystal habit can be tuned from cubic, hexagonal to star-like or pillar shapes
depending on the particle size, shape and magnetic field strength. Further, Wetterskog et al.\,\cite{Nanoscale-2016-8-15571} proposed a phase diagram for the formation of mesocrystals composed of oleat-capped iron oxide nanocubes in a magnetic field. Depending on the strength of the magnetic field and the size, single- or multi-domain crystals and even Rosensweig instabilities are found. In contrast to cubes, evaporation-induced assemblies of $\mathrm{Fe_2O_3}$-nanospheres displayed rhombohedral structures.\,\cite{Scientific-Reports-2017-7-2802} These superstructures rearranged into an fcc-packing later on. This was explained by the capillary pressure at the interface between the saturated and partially saturated regions, which determine the internal order of the superlattices. 

%Dispersed magnetic nanoparticles may be positioned to a specific area by external magnetic fields, allowing for example medical applications. For magnetic nanoparticles, Van-der-Waals attraction and effective dipole-dipole interaction of magnetic nanoparticles can lead to well-ordered assembly structures.\cite{Hindawi} \cite{frontiers-in-china} \cite{SciTechnolAdvMater-15-20142014-055010}
%The assembly of monodisperse magnetic nanoparticles into superstructures is considered as an important step toward fabrication of functional devices. With ongoing miniaturization on these applications, a lower size limit for the magnetic nanoparticles may be reached. In this limit the contributions of magnetic interaction energy, thermal energy and other interaction energies, such as van-der-Waals energy, interfacial interaction energy and steric stabilization energy, may be comparable. Thus, understanding the interplay between particles and assembly structures becomes demanding. 
%The allocation of the different forces leading to the assembly is highly demanding and often still under debate.

Magnetic field induced ordering of nanoparticles is crucial for many applications, where the magnetic particles are dispersed in a liquid. Self-assembled flower-like iron oxide nanoparticles are potential candidates as adsorbent in waste-water-treatment.\,\cite{AdvMater-2006-18-2426} In medical applications, the  coating of the iron oxide nanoparticles might trigger self-assembled structures. Their size plays a significant role in the biocompatibility in human bodies. \cite{SPIONs-ChemRev2008-108-2064} A prominent example in technical applications are ferrofluids.  \cite{Nanofluids-3-1-2012} \cite{J-Magnetics-22-109-2017}   Depending on the external magnetic field and the nanoparticle concentration ferrofluids undergo structural transitions into linear chains along the field direction or more  complex structures such as columns.\,\cite{J-Magnetics-22-109-2017}

\subsection{Crystallization of cuboids}

We describe an in-deep investigation of the self-assembly behaviour of small iron oxides nanocubes in solution (0.2--20\,wt\% in toluene), which  was performed on the oleic acid stabilized 9\,nm sized cubes, depicted in Fig.\,3 %\ref{spheres-cubes-characterization} 
in chapter \ref{synthesis}. The scattering intensities $I(q)$ presented in Fig.\,6
% \ref{cubes2charcterization} 
demonstrate, that the nanocubes are mainly isolated at low concentration (the 2\,wt\% data already shown in Fig.\,3
% \ref{spheres-cubes-characterization}
) and self-assemble at higher concentration (circles). The data at 18\,wt\% show the appearance of additional peaks. Analysis of these Bragg reflections reveals a simple cubic crystal lattice with a unit cell of 14\,nm. This agrees well with face-to-face attachment of 9\,nm sized iron oxide nanocubes, which are covered on all sides by an oleic acid layer, in agreement with the cryo-TEM image of Fig.\,3.
%  \ref{spheres-cubes-characterization}. 
In Fig.\,6 
%\ref{cubes2charcterization} 
on the right hand side an optical microscopic image of the dried sample is presented. It shows three dimensional superstructures in the micrometer range, which we name cuboids in the following. 
\begin{figure} \label{cubes2charcterization}
	\centering
	\includegraphics[width=\figwidth]{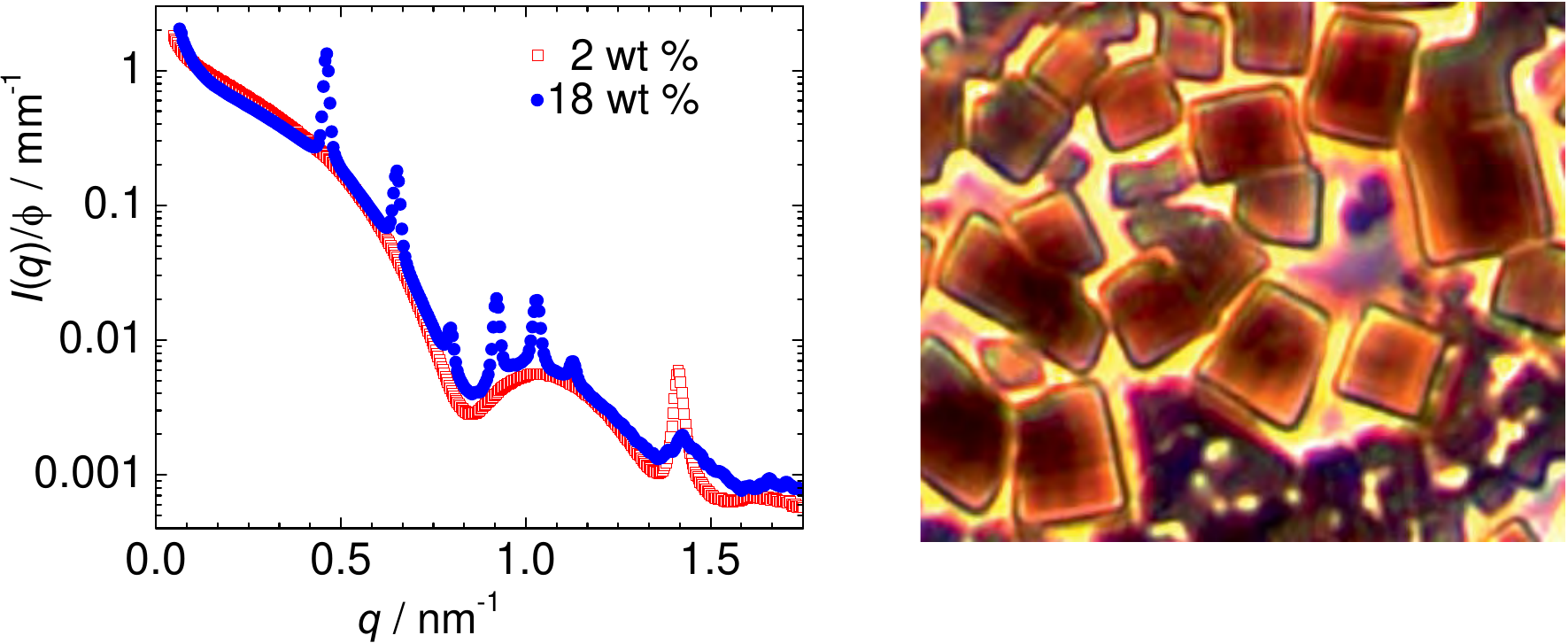}
	\caption{Iron oxide nanocubes in dispersion (SAXS, left) and in dried (optical microscopy, right) state. For 2\,wt\% (open squares) the scattering intensity can be described by isolated cubes, whereas by 18\,wt\% (filles circles) the Bragg reflections clearly indicate a	crystal lattice. The solid sample exhibits large cuboids (micrometer dimensions).  }
\end{figure}
 
\begin{figure}[htb] \label{cubes2charcterization-cryoTEM}
	\centering
	\includegraphics[width=\figwidth]{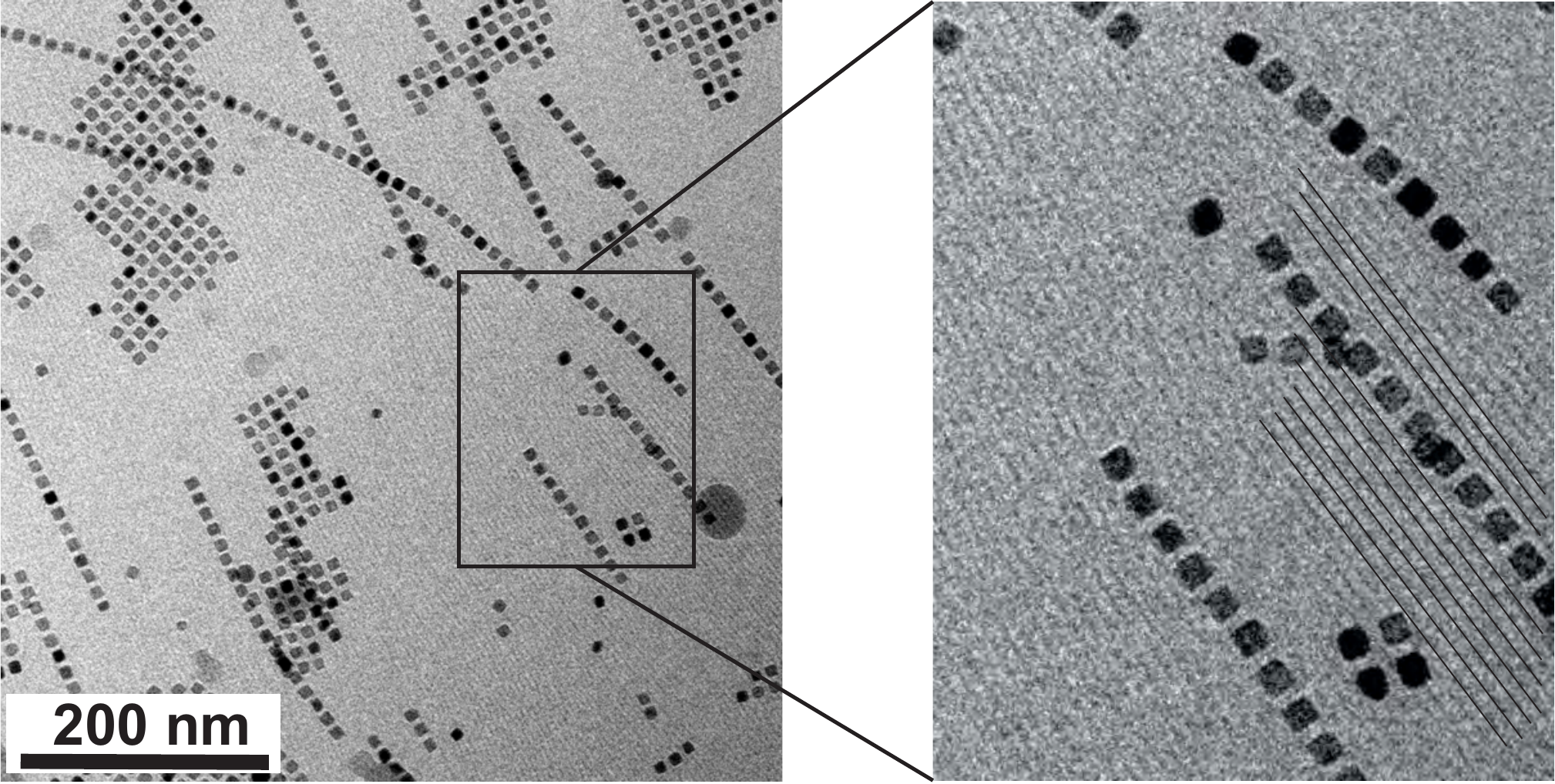}
	\caption{Cryo-TEM-image of the cubic nanoparticles (14\,nm, 0.8\,wt\%) after preparing the sample in a magnetic field (200\,mT). The assemblies clearly reveal oleic acid
layers on the surface of the nanoparticles and free crystalline
oleic acid bilayers. The lines in the zoom-in indicates a co-planar orientation of crystalline oleic acid chains.  }
\end{figure}

In order to understand the assembly of the cubes into cuboids, cryo-TEM studies are performed at lower concentrations of the cubes. A representative example is presented in Fig.\,7,
%\ref{cubes2charcterization-cryoTEM}, 
and more cryo-TEM data are published elsewhere.\,\cite{PNAS-112-14484-2015} They clearly reveal 1D lines and 2D sheets. The lines exhibit a remarkable internal order, namely a face-to-face attachment of the cubes. The sheets show a simple cubic 2d lattice. The zoom-in at the right hand side shows these facts even more clearly. Additionally, a lamellar structure within the solvent (oleic acid and toluene) becomes apparent, as indicated by the thin lines added as a guide to the eye. 
We believe that this structure is formed by the oleic acid chains. The methyl terminals of the oleic acid enables a periodical contrast of the electron density along the lamellar stacking direction. The low density region is attributed to the methyl terminals and the high density region to the dimerizing carboxyl terminals. Similar structures  are reported  for the crystal structure of the $\alpha$--, $\beta$-- and $\gamma$--phase of oleic acid.\,\cite{JAOCS-1985-62-221} \cite{JPhysChemB-1997-101-1803} \cite{Science-2010-329-550} \cite{JPhysChem-1990-94-3180} The lamellar structure of the oleic acid is oriented in the same direction as the 1D structures, which we referred to as lines. Thus, we believe that the crystalline oleic acid bilayers play a crucial role in the stabilization of those lines. In particular it explains the co--planar arrangement of the cube surfaces. 
This explanation is substantiated by the fact, that the self assembly was observed only in samples which showed a significant Bragg reflex caused by crystalline oleic acid (cf. Fig.\,3
% \ref{spheres-cubes-characterization}
). This is underlined further by a similar explanation given by Schliehe et al.\,\cite{Science-2010-329-550} to explain the formation of 2D sheets of PbS-nanocrystals. 

It seems plausible that the crystalline oleic acid bilayers stabilize co--planar structures, when the particles possess plane surfaces. Thus they mediate an oriented attachment of perfect cubes. This agrees with the observation that strongly truncated cubes do not crystallize in simple cubic packing.\,\cite{Nanoscale-2016-8-15571}

\begin{figure} \label{3D-Wuerfel}
	\centering
	\includegraphics[width=\figwidth]{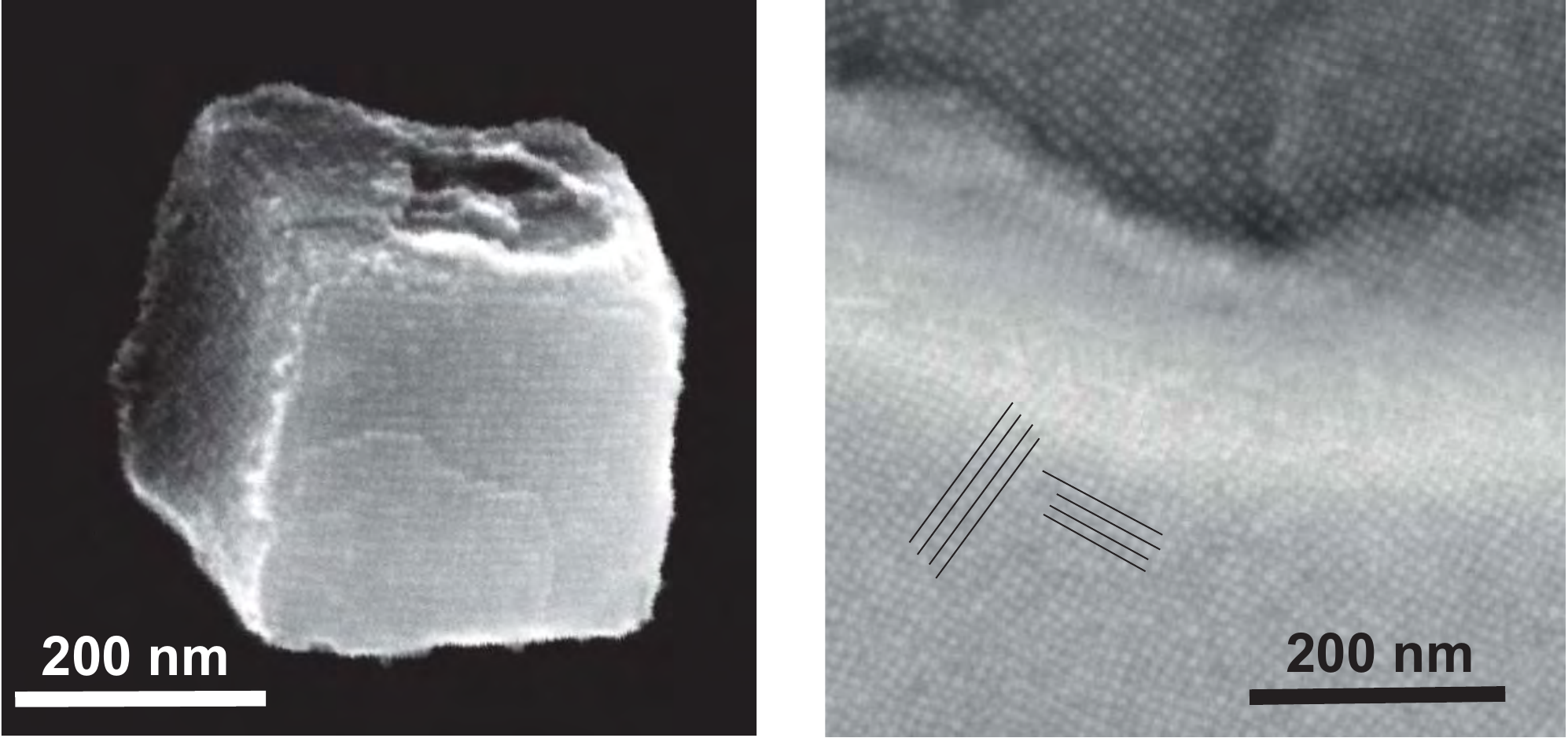}
	\caption{ Self assembly of iron oxide nanocubes to cuboids. (Left) Scanning electron microscopy image of the self-assembled stucture of nanocubes (edge length of 9 nm, 60\% maghemite, 40\% magnetite, oleic acid stabilization).  (Right) SEM image of a zoom into a mesocrystal with micrometer-dimension, consiting of self-assembled iron oxid nanocubes. 
	%The sample is prepared by evaporation drying at standard conditions of a 0.3 wt\% dispersion. 
The black lines are guides for the eyes to visualize the internal order, even near the mesocrystal edge.}
\end{figure}

With this assumption (oriented co-attachment) a solution mediated self-assembly of the iron oxide cubes into 2D sheets  and even 3D cuboids is plausible. Figure 8
%\ref{3D-Wuerfel} 
left hand side shows an example where the nanocubes have self-assembled into a very regular 3D cuboid. This scanning electron microscopic (SEM) image is obtained after drying a 3\,wt\% dispersion in a magnetic field of 130\,mT.
Fig.\,8
% \ref{3D-Wuerfel} 
right presents a zoom into a much larger cuboid  (micrometer range), which was formed in a 0.3\,wt\% dispersion without additional field by evaporation of the solvent toluene at standard conditions. Even this figure demonstrates the astonishing internal order within a cuboidal mesocrystal. The crystal lattice of the nanocubes within both cuboids is best described as simple cubic with defects. It seems that an additional external magnetic field of a 130\,mT is no precondition for the internal sc-ordering of the nanocubes.

Further  SEM images on the dried sample (Fig.\,9)
% \ref{self-assembly-1})
demonstrate  that the formation of perfect cuboids  is not an isolated phenomenon.  Cubes and cuboids with micrometer sizes are found all over the sample. The nearly perfect shapes of the cuboids are clearly seen by the sharp contrasts of the cuboid edges. Looking at the figure one gets the impression that the cuboids may be preferentially  oriented with a longer dimension into the direction of the external magnetic field. We will adress this point in chapter \ref{selfassemblycubeschapter}.
 \begin{figure} \label{self-assembly-1}
 	\centering
 	\includegraphics[width=\figwidth]{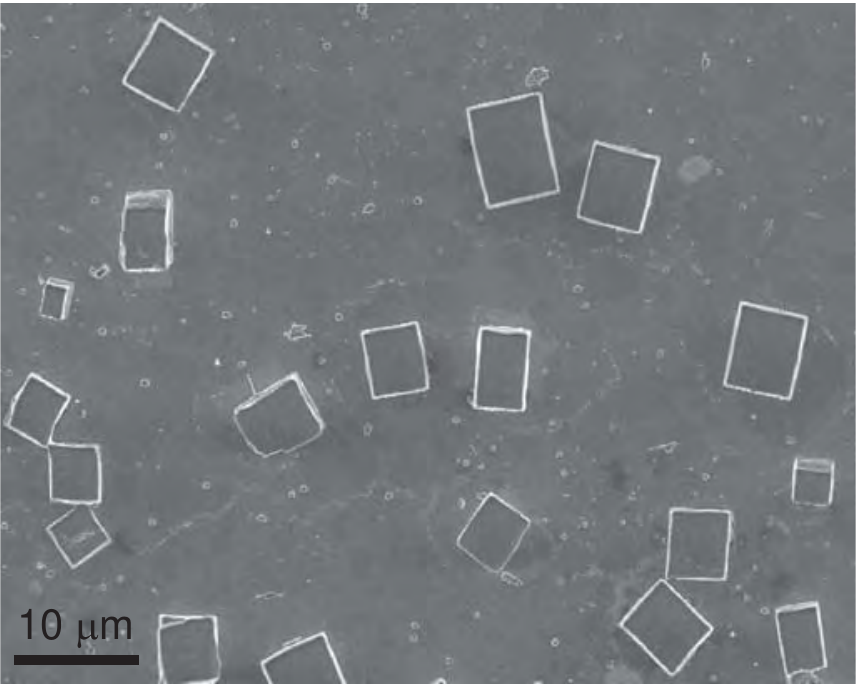}
 	\caption{SEM image of mesocrystalls, consiting of self assembled iron oxid nanocubes (edge length of 9 nm, 60\% maghemite, 40\% magnetite, oleic acid cover). The samples were prepared by evaporation drying at standard conditions of a 0.3\,wt\% dispersion, in presence of a magnetic field of 300\,mT. }.  
 \end{figure}

Additionally performed cryo-SEM experiments (Fig.\,10)
% \ref{cryo-SEM-neu}) 
demonstrate that such large cuboids are already present in dispersion, thereby demonstrating that the  formation of the 3d structures is not induced by the evaporation of the solvent.
\begin{figure} \label{cryo-SEM-neu}
	\centering
	\includegraphics[width=\figwidth]{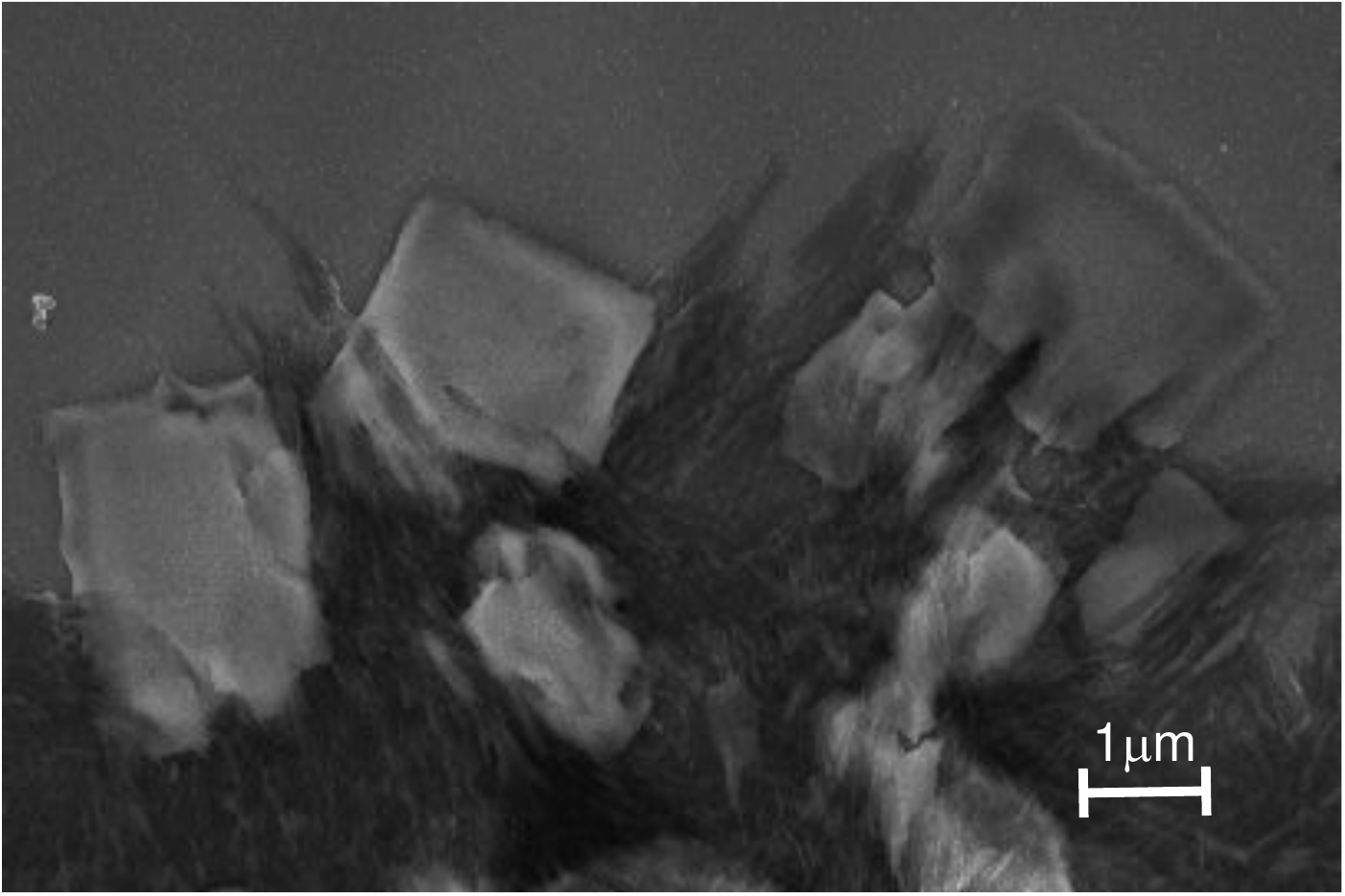}
	\caption{Cryo-SEM image of self assembled cubic nanoparticles (9\,nm iron oxide core, stabilized by oleic acid in toluene). The nanocubes form mesoscaled cuboids, which appear scraggy due to a big amount of overlaying oleic acid. The needle-like dark structures are crystalline oleic acid, which is free in solution. }
\end{figure}

\subsection{Influence of magnetic field}  \label{selfassemblycubeschapter}

The influence of the magnetic field on the self-assembly of iron oxide nanoparticles can be studied indirectly by magnetogranulometry, an experimental me\-thod which makes use of the fact that a suspension of magnetic nanoparticles forms a polarizable fluid. Such a fluid behaves macroscopically like a paramagnetic substance, i.e., it shows no spontaneous magnetization, and no hysteresis in its magnetization curve.  Its susceptibility $\chi _{\mathrm{m}}=\frac{\partial M}{\partial H}$ is concentration and field dependent, and can be orders of magnitude higher compared with normal paramagnets. The magnetization curve $M(H)$ contains information about the size of the nanoparticles dispersed in the fluid --- more precisely, it yields their magnetic moments ($\mu_\mathrm{part}$). The most elementary theoretical description considers a diluted suspension of monodisperse particles with a uniform magnetic moment in an external magnetizing field $H$. The magnetization $M$ should then be given by the Langevin equation:
\begin{math}
\frac{M}{M_{s}}= coth \left( \alpha \right) -\frac{1}{\alpha} \quad \mathrm{with} \quad \alpha = \frac{\mu_\mathrm{part} H}{k_\mathrm{B} T},
\end{math}
the saturation magnetization $M_\mathrm{s}$, the Boltzmann constant $k_\mathrm{B}$, the magnetizing field $H$ and the temperature $T$. If the diluted suspension is polydisperse, a superposition of such Langevin functions can be expected. For higher concentrations of the magnetic particles, both polydispersity and dipole-dipole interaction have to be taken into account, as described by Ivanov et al.\,\cite{Phys-Rev-E-64-041405-2001}\cite{Phys-Rev-E-75-061405-2007}.

Figure 11
%\ref{magnets-1} 
shows 7 examples of such magnetization curves for a 18\,wt\% iron oxide nanocube dispersion obtained with a vibrating sample magnetometer, which measures the magnetic dipole moment by vibrating the sample between a system of pickup coils. A more detailed description of the experimental setup and procedure can be found in Friedrich et al.\ (2012) \cite{RevSciInstr-83-045106-2012}. The first data set was obtained for a fresh sample (crosses, curve at the bottom for $H_i>0$). The sample was subsequently exposed to a magnetizing field of about 800\,kA/m for four hours, and the measurement was repeated ($\times$ symbols). The remaining 5 curves were then obtained after waiting times ranging from 18\,h to 157\,h (filled circles) from the first measurement.

All curves demonstrate paramagnetic behaviour in the sense that no magnetization is observed without a magnetizing field, as expected for a dispersion of particles in a fluid. Having in mind that the initial slope of the Langevin function is determined by the magnetic moment of the particles the ongoing increase in the initial slope of the magnetization curves with time can be qualitatively interpreted: There is an increasing fraction of particles with larger magnetic moment, which is likely to be a direct consequence of the aggregation of the nanoparticles. Indeed, cryo-SEM pictures of the aged sample show larger --- micrometer sized ---  cuboidal objects (Fig.\,10).
% \ref{cryo-SEM-neu}). 
The cuboids appear scraggy due to a big amount of overlaying oleic acid and consist of self-assembled nanocubes. 

\begin{figure} \label{magnets-1}
	\centering
	\includegraphics[width=\figwidth]{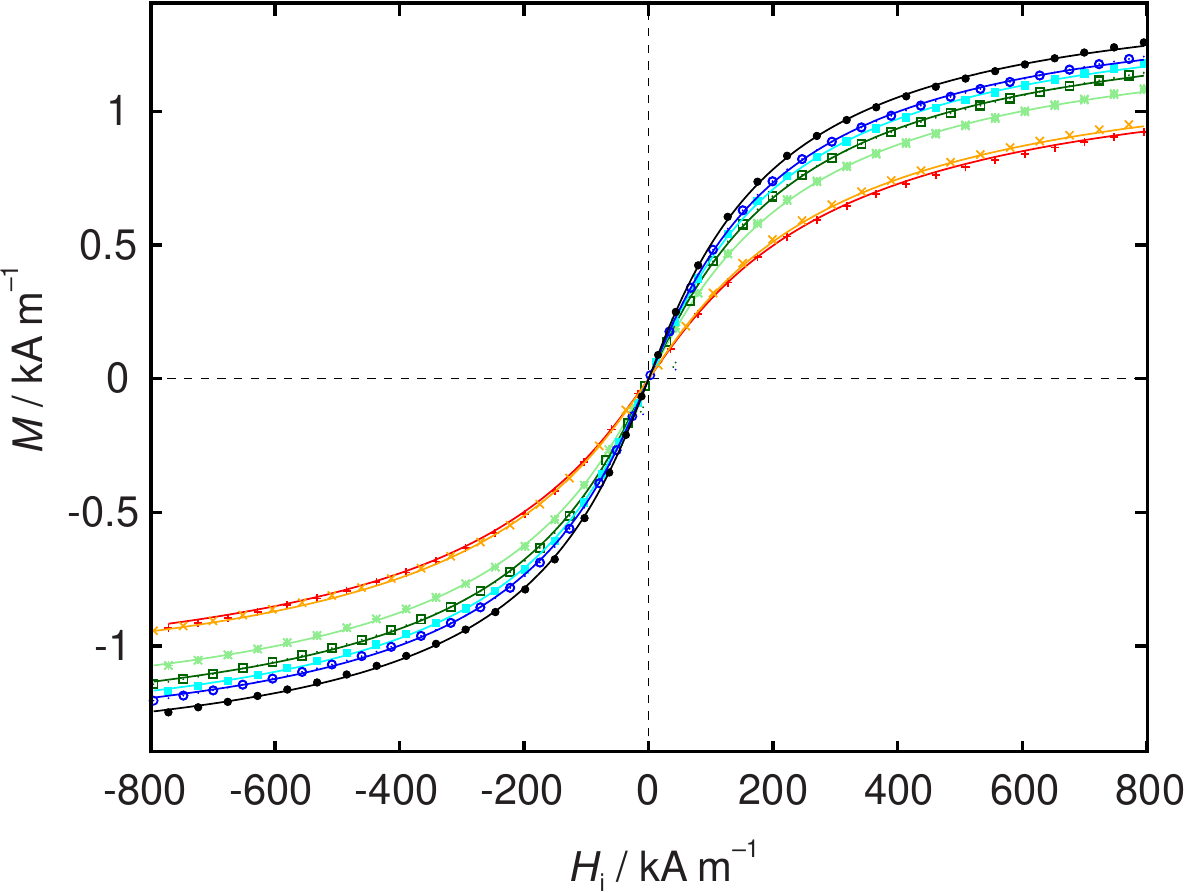}
	\caption{The ageing of cubic nanoparticles (8\,wt\%, iron oxide, edge length 9\,nm) in solution (toluene and oleic acid) triggered by a magnetic field (800\,kA/m for 4\,h) as detected by a vibrating sample magnetometer. The first two data sets are magnetization curves measured before and immediately after that triggering event. The other 5 sets were obtained after waiting times of  18\,h, 40\,h, 65\,h, 89\,h and 157\,h waiting time. During the measurements the magnetizing field strength went from  800\,kA/m to -800\,kA/m and back to 800 kA/m (about $\pm$ 1\,T), which takes about 108\,minutes for one curve.}
\end{figure}

Moreover, the saturation magnetization can be obtained from the measurements, namely by extrapolating the data to very large magnetizing fields. In lowest approximation, a fitted Langevin equation could be used for that. For the data shown in Fig.\,11
%\ref{magnets-1} 
this model showed a systematic deviation from the magnetization curves of the nanocube dispersion. Thus, the refined model from Ivanov et al.\,\cite{Phys-Rev-E-64-041405-2001}\cite{Phys-Rev-E-75-061405-2007} was fitted to the data (solid line). With its additional fit parameter it describes the data perfectly well. Thus we consider it more reliable to extrapolate the data to the saturation magnetization. It turns out that this saturation magnetization increases monotonically with time. In case of iron oxide nanocubes this change continues over days. The saturation magnetization for the 9\,nm iron oxide cubes in toluene turned out to be $M_\mathrm{s}=1.6\,\mathrm{kA/m}$ after one week (Fig.\,11).
% \ref{magnets-1}). 
This value is significantly lower than 100\,kA/m, the theoretically expected value for an iron oxide dispersion of 8\,wt\%. \cite{Rosensweig}
%\textbf{RE Rosenzweig (1997) Ferrohydrodynamics (Dover, Mineola, NY), fig. 2.13}  

The interpretation of the increase of the saturation magnetization  $M_\mathrm{s}$ is less obvious than the interpretation of the initial slope of the curves. In lowest approximation, one would assume that the sum of all magnetic moments of the nanoparticles remains the same, i.e., $M_\mathrm{s}$ should be independent from the amount of clustering. The measurement indicates that this is an over-simplified picture. It seems that particles embedded in an assembly show a stronger magnetic moment than isolated particles. This might be caused by the fact that free surfaces are effectively reduced in an assembly of nanoparticles.

\begin{figure} \label{rehberg-2}
	\centering
	\includegraphics[width=\figwidth]{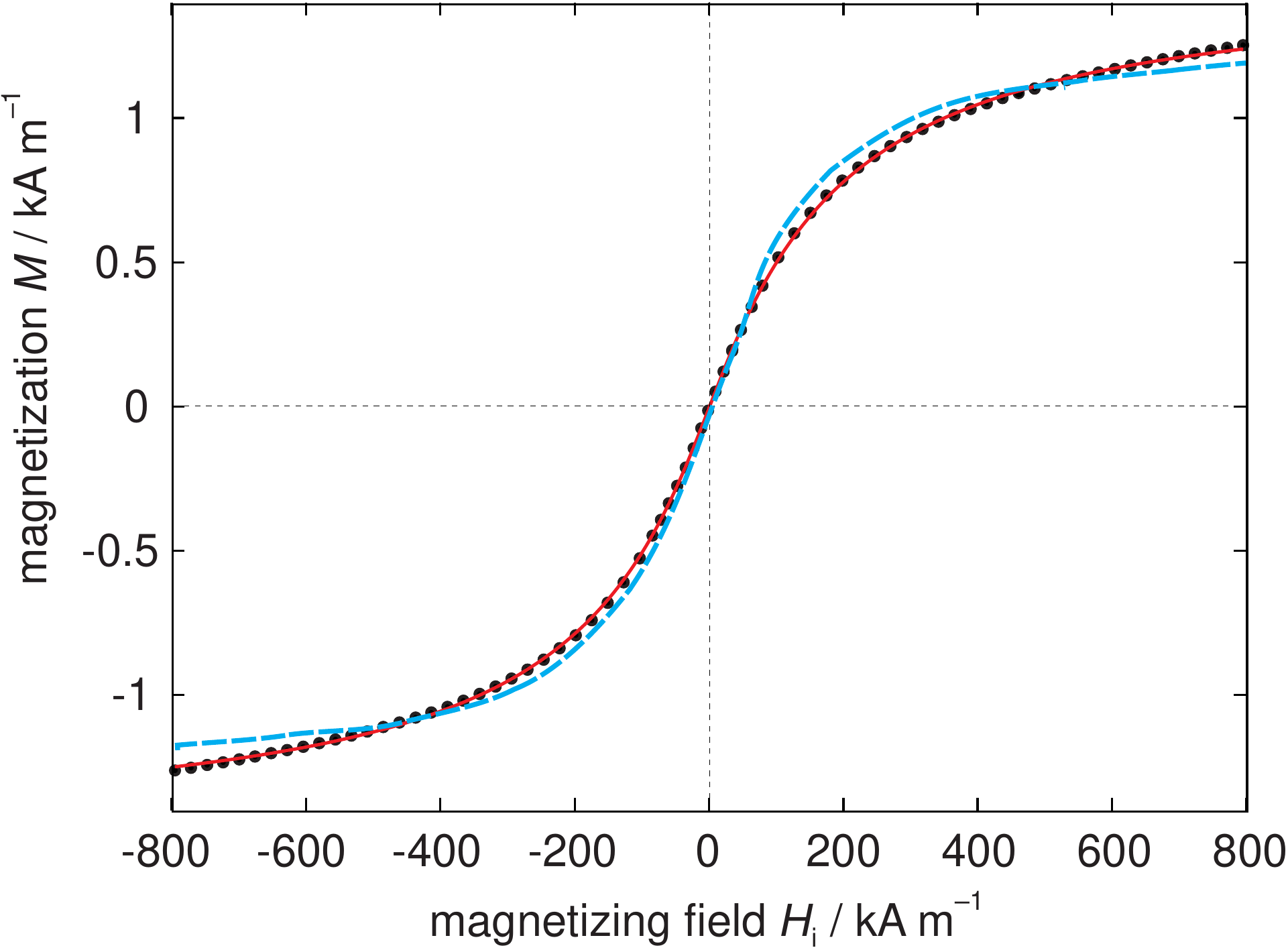}
	\caption{Magnetization of cubic nanoparticles solution as function of the magnetic field strength. The filled circles are the experimental data already shown in Fig.\,11.
%\ref{magnets-1}. 
The dashed line is a fit of a Langevin function for a monodisperse solution, whereas  the solid line represents a fit for a bidisperse solution.}
\end{figure}

Figure 12
%\ref{rehberg-2} 
contains a detailed analysis of the data set from Fig.\,11
%\ref{magnets-1} 
obtained after 157\,h. The measurement exhibits a systematic deviation from the fitted Langevin function (dashed line) expected for a diluted monodisperse solution. This is interpreted as a consequence of the polydispersity of the dispersion, containing a mixture of nanoparticles and cuboidal clusters. When neglecting the dipole-dipole interaction in a dilute solution,  the magnetization curve is expected to be a superposition of Langevin curves, according to the distribution function of the magnetic moments in the polydisperse mixture. In principle, the distribution function of the magnetic moments could be extracted from the curve by the Ivanov fit shown above, but it has to be kept in mind that this method is based on solving a mathematically ill-posed problem. It means, that many different distribution functions will lead to almost the same magnetization curves. 

To circumvent this difficulty and to bring out the essence of cluster formation more clearly, the magnetization data were modelled by assuming that the dispersion would be built from only two different particle sizes, e.g., nanoparticles with a small magnetic moment $\mu_1$ and assemblies with a big magnetic moment $\mu_2$. The moments of the nanoparticles $\mu_1$ and $\mu_2$, as well as the concentration of corresponding particles were used as fit parameters, yielding $\mu_1=1269 \, \mu _\mathrm{B}$ and $\mu_2=7225 \, \mu_\mathrm{B}$  (cf.\ solid line).   

This simplified ansatz describes the experimental data perfectly within the experimental resolution.  Note, that this does not mean that the suspension is really bi-disperse. It rather shows, that the magnetogranulometry reaches its limits at this point --- it cannot convincingly be taken further to obtain a real distribution of the magnetic moments. However, the results of the fitting are still worth to be considered. The solution in this stage should contain 40\,\% particles with a magnetic moment of about 1300 Bohr magnetons, and 60\,\% of particles with a magnetic moment 6 times as big. 
 
To gain insight into the geometrical size of the bigger particles, a connection between magnetic moments and cluster size must be established. In a first attempt to model the magnetization of the clusters, the dipole-dipole interaction of identical magnetic dipoles (the nanocubes) in the arrangement of a simple cubic lattice is considered theoretically. An exact analysis yielding all the the stationary solutions for the freely adjustable dipoles has only been given for the simplest cuboid, containing only $2\times 2\times 2 =8$ nanoparticles. For this geometry the most important dipole arrangement, the ground state, carries no magnetic moment at all \cite{PhysRevB-91-020410-2015}. For bigger clusters, a relaxation code described by Rehberg et al.\ (2015)  \cite{PhysRevE-91-057201-2015} was used to relax dipoles located at fixed positions in simple cubic lattices into some minimum --- not necessarily the ground state --- of the interaction potential. The results are presented in Fig.\,13.
%\ref{rehberg-3}. 
\begin{figure} \label{rehberg-3}
	\centering
	\includegraphics[width=\figwidth]{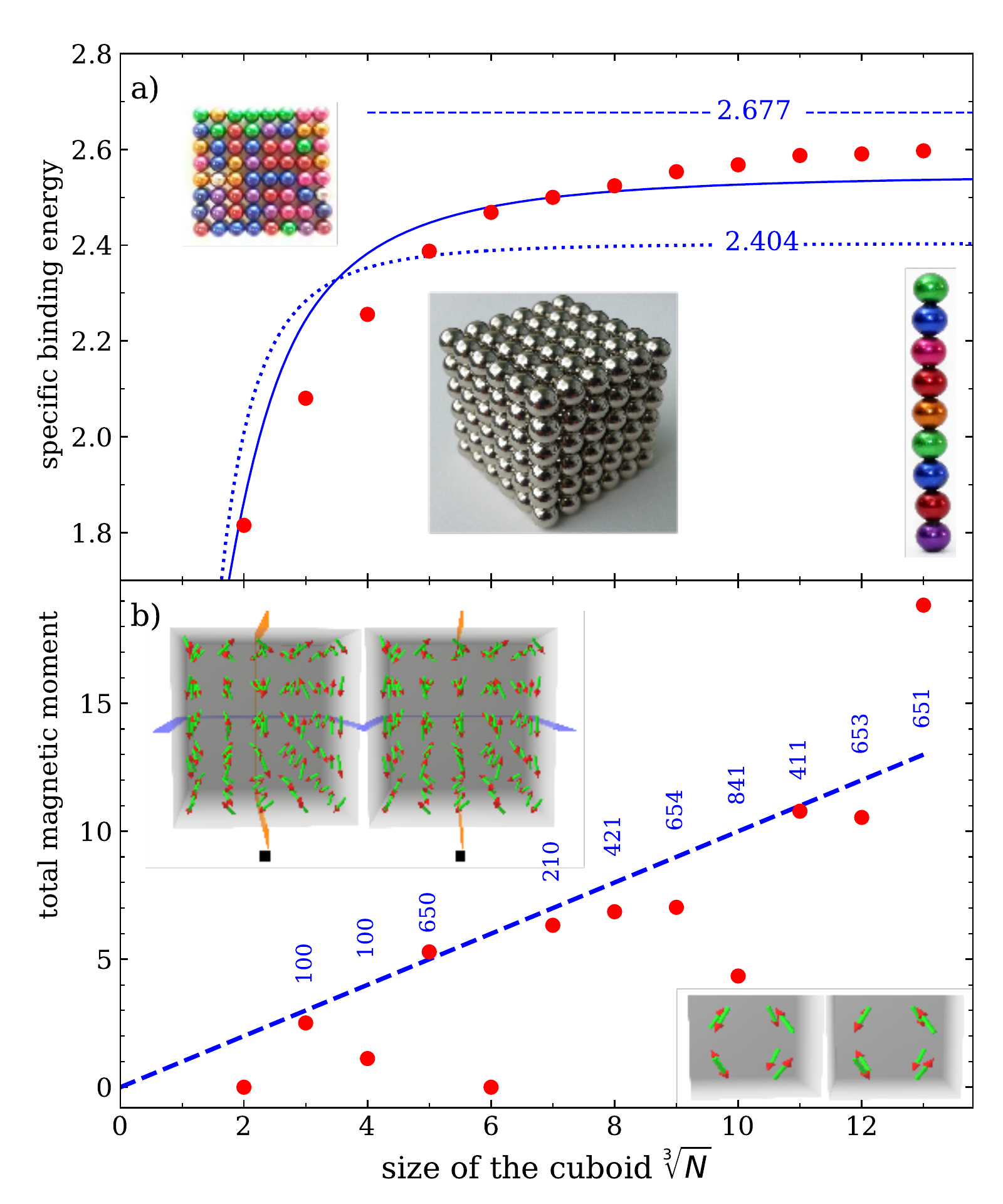}
	\caption{Computationally obtained magnetic moment and specific binding energy of an equilibrium state of freely adjustable dipoles in a cuboidal cluster. The size of the cuboid is given by $\sqrt[3]{N}$, where $N$ is the number of dipoles forming the cluster. a) The specific binding energy of the cuboid is indicated by the filled circles. The inset in the middle shows a photo of a $6\times 6\times 6$ cuboid build from 5\,mm spherical magnets. For comparison, the corresponding energy for dipoles arranged in a 1d line (dotted curve, with the corresponding inset at the right hand side) or a 2d sheet (solid curve, with the corresponding inset at the left hand side) are also indicated. The dashed horizontal line indicates the binding energy in an infinite simple cubic lattice. b) The filled circles indicate the magnitude of the total magnetic moment. The insets show stereographic images of the equilibrium configuration of dipoles for a $5\times 5\times 5$ (left hand side)and a $2\times 2\times 2$ (right) cuboid. The dashed line illustrates the rule of thumb for the magnetic moment. The numbers represent the direction of the magnetization, i.e., $\langle 1,0,0\rangle$ is along the x-axis of the cuboid, and $\langle6,5,4\rangle$ is close to the volume diagonal. No direction is given when the total magnetic moment is 0.}
\end{figure}
 
The upper figure (a) shows the numerically obtained magnetic part of the binding energy per particle within the cluster as a function of the size of the cluster, starting from a $2\times 2\times 2$ cuboid, and going up to $13\times 13\times 13 =2197$ particles. The dimensionless specific binding energy plotted here is scaled by the specific energy that would be needed to disassemble of pair of 2 dipoles, located at a distance of the lattice parameter $a$. To give an example: The value of $2-\sqrt{2}/16-\sqrt{3}/18\approx 1.815$ obtained for the smallest cuboid \cite{PhysRevB-91-020410-2015} means that the energy needed to disassemble this cuboid completely is $8 \cdot 1.815 / 2 \approx 7$ times the energy that would be needed to pull two magnetic dipoles of distance $a$ apart. It turns out that this energy is a monotonically increasing function of the cluster size. It is expected to approach the asymptotic value of about 2.677 indicated by the dashed line, which is the value expected for an infinite simple cubic lattice \cite{Belobrov-1983}. The solid line indicates the energy in the corresponding 2d arrangement, and the dotted line gives the energy of a one-dimensional chain of dipoles with its asymptotic limit 2.404. The insets show macroscopic realisations for the 1d- 2d- and 3d-configuration build from 5\,mm magnetic spheres just for a better illustration of these geometries. So within the family of simple cubic arrangements the rule of thumb would be that for dipole numbers below 30 one-dimensional arrangements are energetically favoured, for intermediate numbers between 30 and 300 a two-dimensional checker-board arrangement maximizes the binding energy. For more particles a cuboid is the most stable arrangement within this sc-family. 
 
The magnetic moment of the equilibrium arrangement of the dipoles in a cuboid is plotted in the lower part (b). The inset provide stereographic images of the equilibrium configuration of dipoles in a $5\times 5\times 5$ and a $2\times 2\times 2$ cuboid.The total moment presented here is scaled with the moment of a single dipole. While the $2\times 2\times 2$ and the $6\times 6\times 6$ cuboid have no magnetic dipole moment at all, the $3\times 3\times 3$ cuboid has a moment of about 3, and the $5\times 5\times 5$ cuboid a moment of 5.  It turns out that in particular the odd-numbered cuboids contain a magnetic moment. This moment increases with the cluster size, and as a rule of thumb one could conclude from this figure that the moment grows approximately as $\sqrt[3]{N}$, which is indicated by the dashed line.  When applying this rule to the interpretation of the fitting result of  Fig.\,12,
%\ref{rehberg-2}, 
the cuboids should be formed from about 200 nanoparticles. 

The orientation of the magnetization shown in the numerical results presented in Fig.\,13
%\ref{rehberg-3} 
does not show a clear tendency for a preferred orientation. This triggers the question about the orientation of the magnetization in real cuboids like the ones shown in Fig.\,9.
%\ref{self-assembly-1}. 
To address this question experimentally, we exposed the cuboids to an external field and recorded the orientation with small angle x-ray scattering (SAXS). The external field was applied either perpendicular or parallel to the incident x-rays. The corresponding scattering patterns $I(q_x,q_y)$ of a 18\,wt\% dispersion in a perpendicular applied magnetic field of 0.1\,mT and 0.98\,T are given in Fig.\,14,
%\ref{magnets-2}, 
and one for a parallel field of 0.79\,T. 
\begin{figure} \label{magnets-2}
	\centering
 	\includegraphics[width=\figwidth]{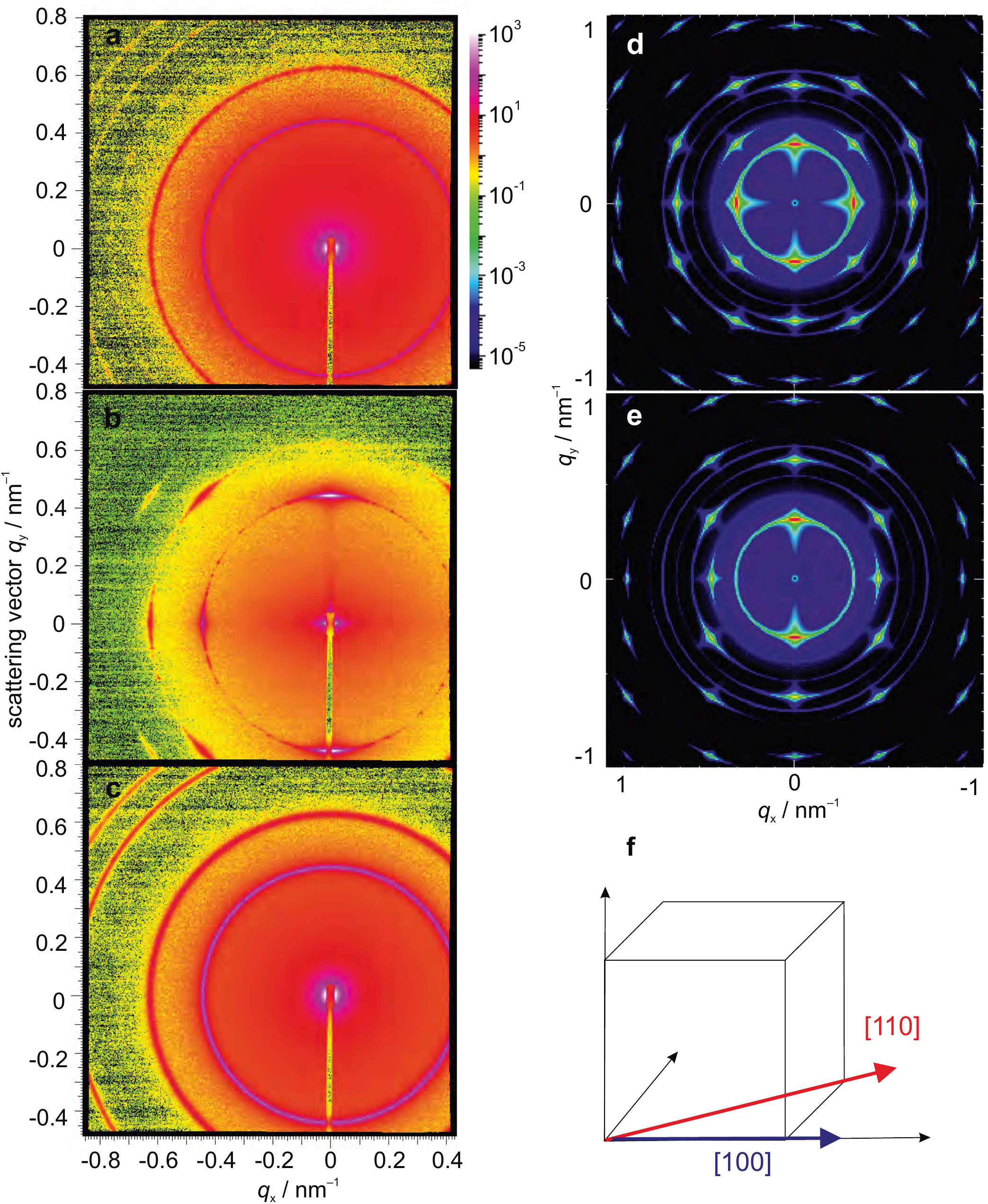}
 	\caption{SAXS patterns of nanocubes. (Left) Synchrotron SAXS-patterns for 18 wt\% iron oxide nanocubes of 9\,nm size in toluene. The data are recorded at earth field (a), within a magnetic field of 0.98\,T oriented perpendicular (b) or of 0.79\,T (c) oriented parallel to the beam. (Right) Numerically obtained scattering patterns. The calculation is based on 9\,nm sized cubes on a simple cubic lattice with a lattice constant of 14\,nm. For one image the x-ray beam was assumed along the [100] direction (d), for the other one along the [110] direction (e). Additionally, the spacial orientation of the cubes is visualized in (f).}
 \end{figure}

At earth field (0.1\,mT) the scattering pattern $I(q_x,q_y)$ is rotational symmetric due to randomly oriented mesocrystrals (Debye-Scherrer rings). From the analysis of the measured intensity, the crystal lattice, the lattice constants and the mean displacement of the nanoparticles from the ideal lattice points can be obtained, provided that the domain size is given. \cite{AdvCollInterfSci163-2011-53} When using a domain size of $2000\,\mathrm{nm}$, the analysis reveals a simple cubic lattice with a lattice constant of $a=14\,\mathrm{nm}$. The lattice constant is in line with cubic iron oxide nanocrystals of 9\,nm coated by an oleic acid layer (cf.\ Fig.\,3).
%\ref{spheres-cubes-characterization}). 
The displacement of the nanocubes is 1\,nm.  
 
In case of Fig.\,14b,
%\ref{magnets-2} b, 
a magnetic field in y-direction perpendicular to the beam direction (z-axis) was applied.  At 0.03\,T the Debye-Scherrer rings become anisotropic, indicating that the orientation of the cuboids --- and accordingly the one of the nanocubes --- are influenced by the external magnetic field.  This anisotropy increases with increasing magnetic field and results in an arc-like pattern, as seen in Fig.\,14b.
%\ref{magnets-2} b.

The peak seen at \textbf{$q_x=0\,\mathrm{nm^{-1}}$} and  \textbf{$q_y=0.44\,\mathrm{nm^{-1}}$}, i.e., (0,0.44), indicates that the cuboids orient their $\langle 100\rangle$ direction along the field. In this case the scattering pattern is expected to be basically a superposition of the two calculated patterns shown in Fig.\,14%\ref{magnets-2}
d and e, because the cuboids are free to rotate around the axis parallel to the field. The peak seen at (0.44,0.44) on the $\langle 110 \rangle$-ring is then in accordance with this interpretation, as well as the one at ($\sqrt{2}\cdot 0.44,0.44$) on the $\langle 111 \rangle$-ring.

Alternatively, one could have assumed that the cuboids would orient their $\langle 110\rangle$-direction or $\langle 111\rangle$-direction along the field. The vanishing intensity at ($0,\sqrt{2}\cdot0.44$) makes it clear that the $\langle 110\rangle$-direction is not the preferred one, while the decreasing intensity at ($0,\sqrt{3}\cdot0.44$) rules $\langle 111\rangle$ out.  

The azimuthal width of a Bragg spot gives a measure for the orientation fluctuations of the polar angle $\psi$. They decrease with increasing strength of the applied field. At 0.98\,T we obtain \textbf{$\psi = 4^\circ$} for the with of the Gauss distribution assumed in the analysing program. \cite{AdvCollInterfSci163-2011-53}   

When the beam passes parallel to the direction of the magnetic field, Debye-Scherrer rings are observed. In this case the $\langle 111 \rangle$-reflection --- which is present at earth field --- is missing. This ring-pattern can be obtained as the radial distribution of the scattering image of Fig.\,14d,
%\ref{magnets-2} d, 
where indeed the $\langle 111 \rangle$-ring is very weak. Thus this observation also supports the interpretation given above: The cuboids orient their $\langle 100\rangle$ direction along the field.

Additionally, an increase in the forward scattering with increasing magnetic field is observed. There are two explanations for this effect: (i) The cuboids grow by attaching further material,  mediated by increasing magnetic field strength. (ii) The cuboids orient more and more along the field lines, hence the beam probes more matter, which is reflected by the increase of the forward scattering. Both effects may take place simultaneously. 

After switching off the external magnetic field no marked difference in the Bragg reflections of the dispersed cuboids was obtained for at least 30\,minutes. This comes as a surprise when considering the Brownian relation time for micrometer sized particles, which should rather be on the order of seconds. To explain these extremely long relaxation times we propose that the cuboids might form form even larger clusters, presumably chain-like ones, along the direction of the field.  

When the concentration of the 9\,nm sized iron oxide cubes is below 5\, wt\%, the scattering at room temperature shows no pronounced Bragg reflections. Hence, the concentration plays also a significant role for the formation of cuboids. 

To investigate the influence of a magnetic field at lower concentrations, a 1\,wt\% dispersion of 9\,nm iron oxide nanocubes was synthesized by carefully avoiding external magnetic fields, in particular magnetic stirrers, or magnets during precipitation of the nanocubes. The sample was split in four parts, two of them were stored at room temperature, and two at -20\,$^\circ$C. At both temperatures one part was exposed to a magnetic field (130\,mT) and the other not. The time-dependent self assembly of those diluted iron oxide nanoparticles was subsequently followed by dynamic light scattering (DLS). Fig.\,15
% \ref{dls} 
shows the resulting intensity correlation functions.  
 
At room temperature ($25\,^\circ$C),  no change in the intensity correlation function is observed within weeks, neither within a magnetic field of 130\,mT nor without one. The correlation function stays compatible with that expected for monodisperse particles, namely a simple exponential decay with a decay time of 9.8\,$\mu$s. This yields a hydrodynamic diameter of 13\,nm, which is larger than than the 9\,nm iron oxide core due to the oleic acid layer.

In Fig.\,15c,
%\ref{dls}\,(c), 
the dispersion kept at $-20\,^\circ$C in the absence of a magnetic field reveals changes of the correlation function with increasing storage time. While the fresh sample shows a purely exponential decay, deviations become prominent after about a week. The reduced slope for larger delay times indicates that the sample is not monodisperse any more. It now contains a fraction of larger particles, most likely aggregates of the nanocubes.  

Figure\,15d
%\ref{dls}\,(d) 
shows that this effect becomes orders of magnitude stronger when the sample is exposed to a magnetic field of 130\,mT at this temperature. Now the correlation function changes significantly within a week. Within the experimental resolution the curve can be fitted by assuming a bidisperse mixture, where the hydrodynamic radius of the larger component is about 2 orders of magnitude larger the that of the smaller one, as indicated by the ratio of the two decay times extracted by the fit. 
For the interpretation of these fitting results the same care must be taken as for the interpretation of the bidisperse fit in the  magnetogranulometry: The fact that a bidisperse distribution of particle sizes fits the measured curve does not proof that this distribution is correct, but it indicates that the analysis can not taken further due to the ill-posed nature of this mathematical problem.

Visual inspection of the sample stored at $-20\,^\circ$C and in presence of an external magnetic field showed a small amount of sedimented particles after about one week. After longer times, the amount of sedimented particles increased only slightly. After 3 weeks, the long time tail of the correlation still revealed assemblies of about 200-400\,nm. All these observations make it clear that two different species are present in the dispersion: (i) small particles, most probably single dispersed cubes, and (ii) larger particles, mainly the crystallized cuboids, which sediment with time.
 
In summary, magnetic fields are helpful for the formation of cuboids, especially at low temperatures and low concentrations. Once created, the cuboids are stable even without a field.  Increasing the concentration facilitates nanoparticle assembly such that cuboids may also form at room temperature without the help of a magnetic field. 
\begin{figure} \label{dls}
	\centering
	\includegraphics[width=\figwidth]{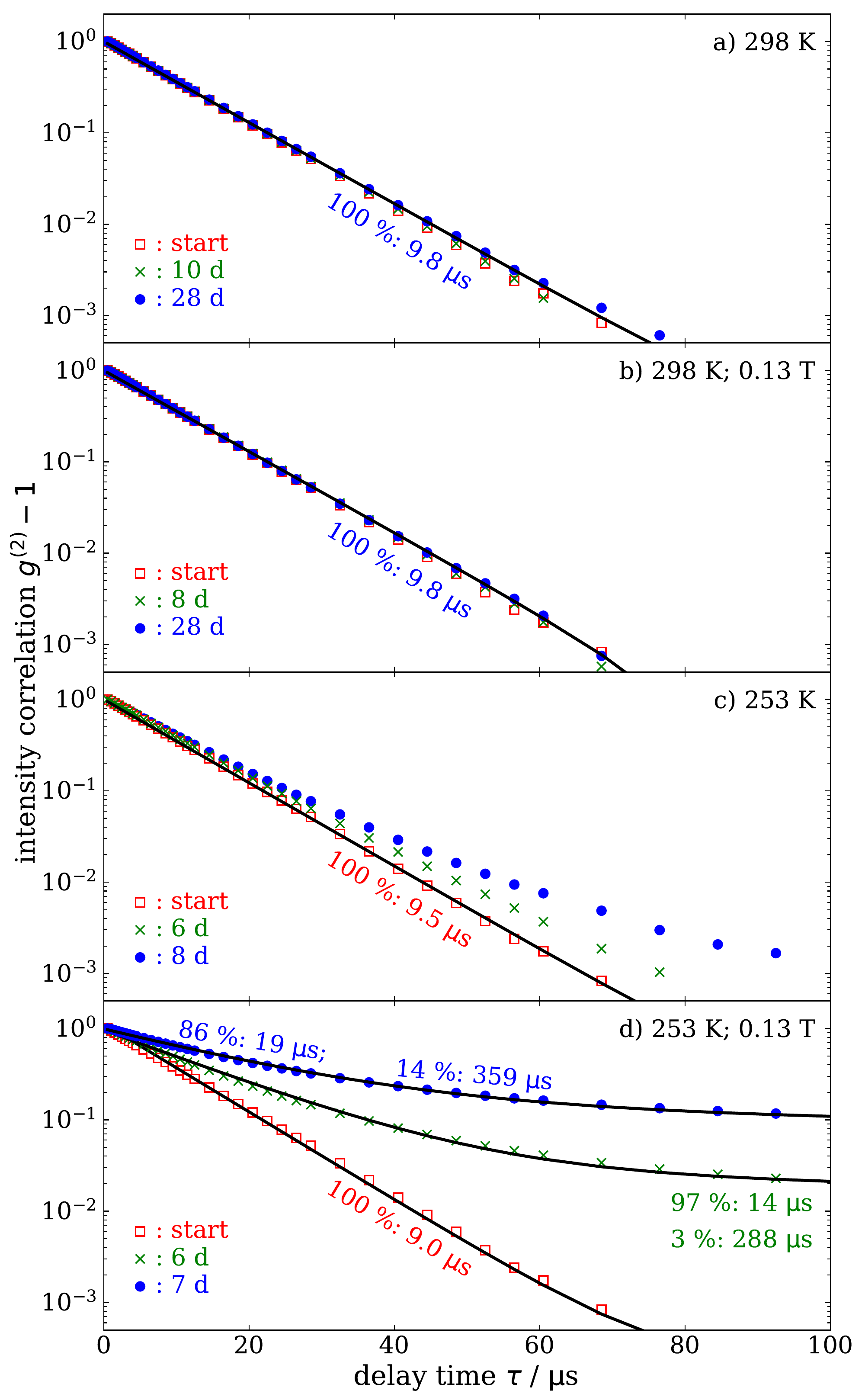}
	\caption{DLS intensity correlation functions $g^2-1$ of a 1\,wt\% dispersions of 9\,nm iron oxide cubes, which are stored under different conditions. (a) Data of the sample stored at $25\,^\circ$C without an additional external magnetic field. The data are taken after 0 (open squares), 10 ($\times$) and 28 (solid circles) days of storage. (b) Data set taken at $25\,^\circ$C after 0 (open squares), 8 ($\times$) and 28 (solid circles) days of storage within a magnetic field of 130\,mT. (c) Intensity correlation functions after 0 (open squares), 6 ($\times$) and 8 (solid circles)days storage of the sample at $-20\,^\circ$C without field. (d) DLS data obtained for the dispersion, which was stored 0 (open squares), 6 ($\times$) and 7 (solid circles) days at $-20\,^\circ$C within a field  of 130 mT. The solid lines indicates fits to bidisperse samples. The relative signal strength and the corresponding decay times of the two components is given by the numbers at the fitted lines.} 
\end{figure}

%\input{T10_conclusion}%T10
%spell-checker britisch 4.11.17, 17 uhr, Ingo

\section{Conclusion}
The understanding of the self assembly of monodisperse magnetic nano\-par\-tic\-les into larger structures like cuboids is a relevant step towards their application, where theses superstructures are either desired (e.g., for the fabrication of functional devices) or are to be avoided (e.g., in biological systems). The formation of self assembled structures strongly depends on the particle shape and size. The self assembly can be followed using (cryo-)TEM, (cryo-)SEM, SAXS, magnetogranulometry, and DLS measurements. These methods reveal that for the cuboids their concentration, the kind of solvent, the temperature and the external magnetic field are additional important parameters. 

The results indicate that the oleic acid used for the stabilization of the nanoparticles has a significant influence on the co-planar orientation of the nanoparticles. The oriented co-attachment of the acid layers simplifies the cuboid formation. 

An external magnetic field helps to form superlattices, but it is not a precondition --- at high concentration cuboids are formed even without the external field. Once formed, they are stable with and without an external field, and can be aligned by such a field.

The stability of cuboids can be explained by considering the surface free energy. When the nanocubes come in close contact up to a distance dictated by the oleic acid layer, they eliminate a pair of high energy surfaces. This energy is larger than the magnetic interaction energy of the nanocubes. This interpretation is substantiated by the fact that no significant indication of self assembly was found for spherical nanoparticles, even at large magnetic fields of 1\,T.

Considering these facts, the kinematics of the self assembly triggered by the magnetic field is an unresolved puzzle. While it is plausible that the field triggers the formation of chains along the field line, the subsequent growth in 2d- or 3d-lattices can not be understood on the basis of magnetic interaction. This force rather seems counterproductive because parallel magnetic chains repel each other. One can speculate that a short magnetic trigger leading to chains, and a subsequent surface energy driven ordering in the absence of an external field might be the most efficient way to built cuboids, a hypothesis that should be tested in forthcoming experiments. 

\section*{Acknowledgement} %spell-checker britisch 4.11.17, 17 uhr, Ingo
The authors thank Peter B\"{o}secke, Markus Drechsler, Beate F\"{o}rster, Andreas Kornowsk, Sara Mehdizadeh Taheri, Theyencheri Narayanan, and Florian M. R\"{o}mer for fruitful cooperation and clarifying discussions.
We thank the German Science Foundation for financial support (DFG-SFB 840). The Bavarian Polymer Institute and the European Synchrotron Radiation Facility (beamline ID2) are gratefully acknowledged for providing measurement times and technical support.

%\begin{thebibliography}{00}
\section*{Literature}
%\input{T11_bibtex_haendisch}%T11
%letzte Aenderung 3.11.,15:31, Ingo

%\bibitem{}
%\end{thebibliography}
\end{document}